\documentclass{article}
\usepackage{fullpage,latexsym,amssymb,stmaryrd,theorem,fleqn}

\newcommand{\oomit}[1]{}

\newtheorem{all}{Proposition}
\newtheorem{lemma}[all]{Lemma}
\newtheorem{proposition}[all]{Proposition}
\newtheorem{theorem}[all]{Theorem}
\newtheorem{corollary}[all]{Corollary}

{\theorembodyfont{\upshape}\newtheorem{remark}[all]{Remark}}
{\theorembodyfont{\upshape}\newtheorem{definition}[all]{Definition}}
{\theorembodyfont{\upshape}}
\newenvironment{proof}{\noindent
{\bf Proof:}}{$\dashv$}
\newenvironment{proof*}{\noindent
{\bf Proof:}}{$\dashv$}

\def\Next{\mathop{\mbox{\raisebox{1.7pt}{$\scriptstyle\bigcirc$}}}}

\newcommand{\Bi}[1][0mu]{\mathord{
  \mbox{\rlap{$\mskip 4.9mu\mskip#1$\relax
    \raisebox{1.5pt}{$\scriptscriptstyle\mathrm{i}$}}}
  \Box}}
\newcommand{\Di}[1][0mu]{\mathord{
  \mbox{\rlap{$\mskip 5.0mu\mskip#1$\relax
    \raisebox{1.6pt}{$\scriptscriptstyle\mathrm{i}$}}}
  \Diamond}}
\newcommand{\Prev}[1][0mu]{\mathord{
  \mbox{\rlap{$\mskip 3.8mu\mskip#1$\relax
    \raisebox{1.2pt}{$-$}}}
  \Next}}

\def\Fin{\mathit{fin}\,}

\def\NL{\ifmmode \textrm{NL}\else $\textrm{NL}$\fi}
\def\NLone{\ifmmode \NL^{\!1}\else $\NL^{\!1}$\fi}
\newcommand{\QLTL}{\ifmmode \textrm{QLTL}\else $\textrm{QLTL}$\fi}
\newcommand{\QPITL}{\ifmmode \textrm{QPITL}\else $\textrm{QPITL}$\fi}
\newcommand{\PITL}{\ifmmode \textrm{PITL}\else $\textrm{PITL}$\fi}
\newcommand{\CPL}{\ifmmode \textrm{CPL}\else $\textrm{CPL}$\fi}

\newcommand{\existsMN}[2]{{{\exists}_{#1}^{#2}}}

\def\midd {{\,\mid\,}}

\def\PiInv{{\Pi^{-1}}}

\def\True{\mathit{true}}
\def\False{\mathit{false}}
 
\def\Imp{\supset}
\def\Skip{\mathit{skip}}
\def\Empty{\mathit{empty}}

\def\Inf{\mathit{inf}}
\def\Finite{\mathit{finite}}

\let\defeq=\triangleq
\def\Var{\mathrm{Var}}

\def\pref{\mathsf{pref}}

\newcommand{\angles}[1]{\langle#1\rangle}
\newcommand{\Cl}{{\mathit{Cl}}}

\newcommand{\dom}{{\mathrm{dom}\,}}
\newcommand{\ITL}{\ifmmode \textrm{ITL}\else $\textrm{ITL}$\fi}
\newcommand{\ITLNL}{\ifmmode \textrm{ITL{-}NL}\else $\textrm{ITL{-}NL}$\fi}
\newcommand{\LTL}{\ifmmode \textrm{LTL}\else $\textrm{LTL}$\fi}
\newcommand{\HS}{\ifmmode \textrm{HS}\else $\textrm{HS}$\fi}
\newcommand{\DC}{\ifmmode \textrm{DC}\else $\textrm{DC}$\fi}
\newcommand{\PLTL}{{\mathrm{PLTL}}}

\newcommand{\tr}[1]{{\mathsf{t}\left({#1}\right)}}

\font\hollow=msbm10

\newcommand{\Zset}{{\mbox{\hollow Z}}}

\newcommand{\parag}[1]{\noindent{\bf #1}\ }

\let\D=\displaystyle

{\makeatletter
\global\let\myignoretrue=\@ignoretrue
}

\begin{document}

\title{Applications of Interval-based Temporal Separation: the Reactivity Normal Form, Inverse $\Pi$, Craig Interpolation and Beth Definability}

\author{
Dimitar P. Guelev\\
Institute of Mathematics and Informatics\\
Bulgarian Academy of Sciences\\
E-mail: {\tt gelevdp@math.bas.bg}
}

\maketitle

\begin{abstract}
We show how interval-based temporal separation on the extension of Moszkowski's discrete time interval temporal logic (Moszkowski, 1986) by the neighbourhood modalities ($\ITLNL$) and a lemma which is key in establishing this form of separation in (Guelev and Moszkowski, 2022) can be used to obtain concise proofs of an interval-based form of the reactivity normal form as known from (Manna and Pnueli, 1990), a new normal form for $\ITL$ formulas which, given a state formula $w$, features the conditions that the maximal $\Box w$- and $\Box\neg w$-subintervals of an interval satisfying the given formula need to satisfy, the expressibility of the inverse of the temporal projection operator from (Halpern, Manna and Moszkowski, 1983) in $\ITL$, the elimination of propositional quantification in $\ITLNL$ and, consequently, uniform Craig interpolation and Beth definability for $\ITLNL$.

\noindent
{\bf keywords:}
interval temporal logic $\cdot$ Gabbay separation $\cdot$ temporal reactivity temporal form $\cdot$ temporal projection $\cdot$ Craig interpolation $\cdot$ Beth definability.
\end{abstract}

\section*{Introduction}
\label{introduction-sec}

In \cite{GM2022} we established a separation theorem for the extension of Moszkowski's discrete time $\ITL$ Interval Temporal Logic ($\ITL$, \cite{DBLP:conf/icalp/HalpernMM83,Mos85,Mos86,CMZ}) by the {\em neighbourhood modalities}, which are known from Halpern-Shoham logic ($\HS$, \cite{HS86}) and Neighbourhood Logics ($\NL$s). The latter are systems of interval-based temporal logic with these modalities as the only temporal operators, cf. e.g., \cite{DBLP:journals/eatcs/MonicaGMS11}. The Kripke semantics of the neighbourhood modalities, written $\angles{A}$ and $\angles{\overline{A}}$ in $\HS$, is based on Allen's system of binary relations on time intervals \cite{DBLP:journals/cacm/Allen83}. Our separation theorem was formulated on the model of Gabbay's separation theorem for Linear Temporal Logic with past $\PLTL$ \cite{Gab89}. We have since ported the result to a subset of the Duration Calculus (DC) \cite{DBLP:conf/time/Guelev22}, which is an interval-based real-time logic based on a related \emph{chop} operator \cite{ZHR91,DBLP:series/eatcs/ChaochenH04}. 
In \cite{GM2024} we applied separation to establish the expressive completeness of $\ITLNL$, the considered extension of $\ITL$, wrt the monadic theory of $\angles{\Zset,<}$. This is the first order theory for $\ITLNL$ with just $\Next$, \emph{chop} and the neighbourhood modalities as the temporal connectives, and the second order theory, if \emph{chop-star}, the iterative form of \emph{chop} of $\ITL$, is included. The expressive completeness proof in \cite{GM2024} is established on the model of the proof of expressive completeness of $\PLTL$ from \cite{GRH94}, which is a major improvement on the proof of this seminal result of Kamp \cite{Kamp68}. The expressive completeness of the subset of DC from \cite{DBLP:conf/time/Guelev22} was established in \cite{DBLP:journals/iandc/Rabinovich00} well before separation was established, by another method.

In this paper we show how separation helps establish some more results in $\ITLNL$, whose variants in point-based discrete time propositional temporal logics are known for their many uses. This includes the normal form for reactivity properties from the hierarchy of temporal properties of Manna and Pnueli \cite{MP89a}, the companion theorems of Craig interpolation and Beth definability, and the expressibility in $\ITL$ of the inverse of the projection operator $\Pi$ from \cite{DBLP:conf/icalp/HalpernMM83} wrt its second operand. 

{\em The hierarchy of Manna and Pnueli in $\PLTL$} is the inclusion ordering of a finite number of classes of $\omega$-languages also termed {\em (temporal) properties}. In ascending order, the lowest classes are known as {\em satefy} and its dual {\em guarantee}, aka {\em reachability}. They are followed by the intermediate class of {\em obligation} properties, which are finite combinations of safety and guarantee ones by $\cup$ and $\cap$. Then come the classes of {\em recurrence}, aka {\em response}, and {\em persistence} properties, which are about an event occurring infinitely many times or ultimately always, respectively. The top class, which is called {\em reactivity}, consists of the finitary combinations of recussence and persistence by $\cup$ and $\cap$. In \cite{MP89a}, the hierarchy is given in the notation of topology, involving up to $G_\delta$-sets and $F_\sigma$-sets in the topological space of infinite sequences, the notation of regular $\omega$-languages, and $\PLTL$. Properties from each of the classes that are definable by regular expressions or $\PLTL$ formulas are shown to admit normal forms in these languages. Further insight on the study can be found in \cite{SafetyProgress10.1007/978-3-642-58041-3_5}, where earlier results about the individual classes such as \cite{DBLP:journals/mst/Landweber69,DBLP:journals/iandc/Wagner79} on the definability of recurrence and reactivity by their corresponding types of $\omega$-automata have been both referred to and included for the sake of self-containedness. In this paper we use interval-based separation to establish an interval-based normal form for the top class of {\em reactivity} in $\ITLNL$. The scope of this normal form is greater than that of the $\PLTL$ ones and equals the scope of the regular expression normal form because the expressiveness $\ITLNL$ matches regular $\omega$-languages.

{\em The $\Pi$ operator} was introduced in \cite{DBLP:conf/icalp/HalpernMM83}. Given a state (modality-free) formula $w$ and temporal formulas $A$, $w\Pi A$ means that $A$ is true in the interval consisting of the states of the reference interval which satisfy $w$. $\Pi$ has proven to have multiple applications, including some we made in \cite{DBLP:journals/fac/MoszkowskiG17}. This application stems from the fact that interleaving concurrency is about partitioning processing time between concurrent processes, and that the first operand $w$ of $w\Pi\ldots$ can be used to select the time allocated to a process. It is closely related to reasoning about concurrency in terms of {\em clocks} \cite{DBLP:conf/icalp/EisnerFHMC03}, see also the recent work \cite{DBLP:journals/fac/ZhangMZL24}.
$\Pi$ is known to not increase the ultimate expressive power of $\ITL$: every formula in $\ITL$ with $\Pi$ has an equivalent one without, and the latter can be obtained by a compositional translation. In terms of our application to concurrency, that translation can be seen as giving an explicit and therefore much more verbose and lower-level account of how interleaving works. 

In this paper we introduce the inverse $\PiInv$ of $\Pi$ wrt its second operand $A$, while assuming the first operand to be fixed. Intuitively, $\sigma\models w\PiInv A$ means that $\sigma$, which is supposed to satisfy $w$ in all along, can be expanded by inserting non-$w$ states to obtain an interval which satisfies $A$. This can be seen as determining whether the reference interval $\sigma$ consists of the `observable' $w$-states of a possibly longer interval that accommodates a process which fits the description $A$. We use separation to show that $\PiInv$ is expressible in $\ITL$ too. 

Along with the expressibility of $\PiInv$, we use separation to obtain a new normal form for $\ITL$ formulas. We find this normal form interesting because, given an arbitrary formula $A$, it allows reading the conditions for an interval $\sigma$ to satisfy $A$ in terms of the properties of the subintervals of $\sigma$ that are maximal either wrt the condition of satisfying some state formula $w$ all along, or the condition of satisfying $\neg w$ all along. Partitioning time in this way was proposed for the analysis of hybrid systems in \cite{PD97}. In that work time was proposed to be viewed as an alternating sequence of {\em macro}-time intervals for modeling continuous processes and {\em micro}-time intervals for the computation steps of a digital controller. This facilitated ignoring the duration of micro-time for the sake of simplicity while still maintaining the causal ordering of computation steps.

A logic has the {\em Craig interpolation property} if, for any valid implication $A\Imp B$, there exists an {\em interpolating} formula $C$ built using only the non-logical symbols that appear in both $A$ and $B$ such that $A\Imp C$ and $C\Imp B$ are valid too. First order predicate logic is known to have this property. Despite being expressively equivalent to a monadic first-order theory, propositional Linear Temporal Logic ($\LTL$, cf. e.g., \cite{Pnu77}) does not have Craig interpolation \cite{Maksimova91} but some subsets of $\LTL$ have it \cite{DBLP:conf/csl/GheerbrantC09}. $\ITL$ and $\ITLNL$ have these theorems thanks to the possibility to express propositional quantification using \emph{chop-star}. Since propositional quantification enables writing {\em strongest consequences}, which are also uniform interpolants, temporal logics with point-based Kripke semantics where propositional quantification is either present or expressible as in, e.g., the propositional $\mu$-calculus \cite{Koz83,DBLP:conf/csl/GheerbrantC09}, have the properties too. The linear time $\mu$-calculus (cf. e.g., \cite{DBLP:conf/concur/Kaivola95}) and $\ITL$ have it too thanks to the expressibility of propositional quantification. Interpolants $C$ can be derived from just one of the given $A$ and $B$ in these logics. This is the reason for calling this stronger form of interpolation {\em uniform}. (To the best of our knowledge, first-order predicate logic does not have uniform interpolation.) 

Craig interpolation is closely related to two more model-theoretic properties: {\em Robinson consistency} and {\em explicit (Beth) definability}, both enjoyed by first order predicate logic, cf. e.g., \cite{Sho67}. Once a logic is shown to have one of the three properties, the other two often follow by standard arguments. Let $[p'/p]A$ stand for the substitution of $p'$ for $p$ in $A$. Sentence $A$ {\em implicitly defines} non-logical symbol $p$, which is assumed to appear in $A$, if, given some fresh symbol $p'$ of the type of $p$, $A\wedge [p'/p]A$ implies that $p$ and $p'$ are semantically equivalent. For $p$ being a predicate letter in first order predicate logic this can be written as
\begin{equation}\label{bethCPL}
\models A\wedge [p'/p]A\Imp (p(\overline{x})\equiv p'(\overline{x}))
\end{equation}
where $\overline{x}$ is a sequence of individual variables. Beth's work shows that in first order predicate logic {\em implicit} definability implies the existence of a predicate formula $C$ with $\overline{x}$ as the list of free variables such that
\[\models A\Imp(p(\overline{x})\equiv C),\]
that is, the implicit definability of $p$ implies the existence of a $C$ which {\em explicitly} defines $p$.

Beth definability can be formulated in $\LTL$ as follows. Propositional variable $p$ is {\em implicitly defined by formula $A$}, if
\begin{equation}\label{bethLTL}
\models\Box A\wedge \Box [p'/p]A\Imp \Box(p\equiv p').
\end{equation}
The $\Box$ of $\Box A$ corresponds to the requirement on $A$ of the predicate logic case to be a sentence. The $\Box$ of $ \Box(p\equiv p')$ states that $p$ and $p'$ are equivalent at all times, that is, they stand for the same unary predicate on time. The corresponding first order case $\forall \overline{x}$ is omitted because it can be derived by generalization. The $\Box$ of $ \Box(p\equiv p')$ can be omitted for the same reason too. The counterexample from \cite{Maksimova91} indicates that some $p$ which `count' can be implicitly defined, whereas `counting' is known to be not expressible in $\LTL$ (cf. e.g. \cite{Zuc86}); hence a formula $C$ such that $\models \Box A\Imp\Box(p\equiv C)$ is not bound to exist. By contraposition, the companion Craig property fails for $\LTL$ too.

To establish Craig interpolation in $\ITLNL$, we first use separation to show how the elimination of propositional quantification extends from $\ITL$ to include the neighbourhood modalities. Along with the elimination of propositional quantification in general in the whole of $\ITLNL$, we prove that \emph{chop-star}-free $\ITLNL$ admits the elimination of propositional quantification with an upper bound on the number of the states in which the quantified variable is true.

In $\ITL$, formulas generally define properties of sequences of consecutive states, which are called intervals, but atomic propositions are restricted to depend only on the initial state of the reference interval. This convention is known as the {\em locality principle}. It bridges the gap between $\ITL$'s interval-based semantics and state-based models such as sequences of observations, and Kripke models. $\ITL$'s temporal connectives $\Next$, \emph{chop} and \emph{chop-star} provide reference only to the subintervals of the reference interval. To indicate this, these connectives are called {\em introspective}, unlike many of the connectives in Halpern-Shoham logic ($\HS$, \cite{HS86}), which provide access outside the reference interval and are therefore called {\em expanding}. In introspective $\ITL$, $\Box$ stands for {\em in all suffix subintervals}, and the {\em universal modality} \cite{GorankoPassy1992} can be defined by combining $\Box$ with $\Bi A\defeq\neg(\neg A;\True)$, which means $A$ {\em in all prefix subintervals.}

The locality principle restricts the class of properties which an atomic proposition $p$ can be defined to express explicitly by putting $\Box\Bi(p\equiv C)$ for a temporal $C$ with introspective connectives only to properties of the initial states of reference intervals (which are also degenerate $1$-state intervals). The use of introspective connectives is vacuous in such intervals because they are their only own subintervals: e.g., $C';C''$ simplifies to $C'\wedge C''$. Hence the scope of Beth's theorem is very limited in $\ITL$ without expanding modalities. The neighbourhood modalities are expanding. With these modalities allowed in $\ITLNL$ explicit definitions $p\equiv C$, the expressive completeness of $\ITLNL$ implies that atomic propositions $p$ can be defined to stand for (the $\ITLNL$-translation of) any unary predicate that is definable in $MSO(\Zset,<)$. 

As mentioned above, systems based on $\angles{A}$ and/or $\angles{\overline{A}}$ only are known as Neighbourhood Logics ($\NL$s). Surveys on interval-based temporal logics, including $\NL$s, can be found in \cite{DBLP:journals/jancl/GorankoMS04,DBLP:journals/eatcs/MonicaGMS11}. Importantly \emph{chop}, which is binary, and is basic in systems such as Venema's CDT \cite{DBLP:journals/logcom/Venema91}, Signed Interval Logic \cite{DBLP:conf/csl/Rasmussen99}, Moszkowski's $\ITL$ and its extension by the neighbourhood modalities in this paper, is not expressible in propositional $\HS$ and $\NL$. Together with \emph{chop-star}, also available in $\ITL$, \emph{chop} is key to the result in this paper. Hence the results in this paper, simple to establish by means of separation as they are, assert the choice to focus on $\ITLNL$ as we did by establishing separation after Gabbay \cite{Gab89} and expressive completeness of the extension of $\ITL$ by the neighbourhood modalities in \cite{GM2022,GM2024}. The use of the real-time form of this set of modalities was proposed in \cite{DBLP:conf/compos/ChaochenH97}.

\qquad 

\parag{Structure of the paper:} Section \ref{preliminaries} overviews the known notions and results that are used in the paper, including propositional $\ITLNL$, the extension of $\ITL$ by the neighbourhood modalities, and separation for $\ITLNL$ as known from \cite{GM2022}. The following sections are dedicated to the interval-based reactivity normal form in $\ITLNL$, the normal form for $\ITL$ formulas that highlights the maximal subintervals satisfying some given state condition, inverse $\Pi$, the expressibility of propositional quantification in $\ITLNL$, the expressibility of propositional quantification with an upper bound on the number of states which satisfy the quantified variable in \emph{chop-star}-free $\ITLNL$ and, as immediate corollaries, uniform Craig interpolation and Beth definability in $\ITLNL$.

\section{Preliminaries}
\label{preliminaries}

\parag{Reactivity as in the hierarchy of temporal properties from \cite{MP89a}}
Given a statespace $\Sigma$, a discrete time temporal {\em property} is a set of infinite sequences $\sigma:\omega\rightarrow\Sigma$. Such sequences are also called $\omega$-words; properties are also called $\omega$-languages over $\Sigma$ as the alphabet. The set of all $\omega$-words over $\Sigma$ is written $\Sigma^\omega$. In general ${}^\omega$ is a unary operation from finite word languages to $\omega$-ones akin to Kleene star ${}^*$. Given $L\subseteq\Sigma^+$,
\[L^\omega\defeq\{\sigma_0\cdot\sigma_1\cdot\ldots:\sigma_k\in L,k<\omega\}.\] 
{\em Regularity} as known about finite word formal languages has been generalized to $\omega$-languages (cf. e.g. \cite{DBLP:books/el/leeuwen90/Thomas90}) and has since become a cornerstone in the verification of reactive systems:
$\omega$-Language $L\subseteq\Sigma^\omega$ is {\em regular}, if it admits the form $\bigcup\limits_{k=1}^K L_k\cdot (L_k')^\omega$ for some regular $L_1,\ldots,L_K,L_1',\ldots,L_K'\subseteq\Sigma^+$. 

{\em Reactivity} is the top class of the hierarchy of temporal properties which was proposed in \cite{MP89a}. Given a $\sigma\in\Sigma^\omega$,
\[\pref(\sigma)\defeq\{\sigma^0\ldots\sigma^k:k<\omega\}\]
is the set of $\sigma$'s (finite) prefixes. The properties from 
that hierarchy are defined in terms of $\pref(.)$ on the $\omega$-words $\sigma$ of the respective languages. $L\subseteq\Sigma^\omega$ is a {\em reactivity} property, if 
\begin{equation}\label{reactReg}
\begin{array}{p{5in}}
there exist some $N<\omega$ and $M_n',M_n''\subseteq\Sigma^+$, $n=1,\ldots,N$, such that $\sigma\in L$ if, for some $n\in\{1,\ldots,N\}$, $\pref(\sigma)\setminus M_n'$ is finite and $\pref(\sigma)\cap M_n''$ is infinite. 
\end{array}
\end{equation}
In case $M_n',M_n''$ are regular, this definition implies that $L$ is regular too. Using the De Morgan laws and the closedness of the class of all regular $\omega$-languages and the class of all regular (finite word) languages under complementation, the above definition can be formulated as a conjunction on conditions on some finite number of (some other) pairs $M_n',M_n''\subseteq\Sigma^+$ as follows:
\begin{equation}\label{reactRegConj}
\begin{array}{p{5in}}
There exist some finite number of pairs $M_n',M_n''\subseteq\Sigma^+$, $n=1,\ldots,N$, such that $\sigma\in L$ if, for {\em all} $n\in\{1,\ldots,N\}$ such that $\pref(\sigma)\cap M_n''$ is finite, $\pref(\sigma)\cap M_n'$ is finite too. 
\end{array}
\end{equation}
In the notation of formal languages, regular properties of the various classes from the hierarchy admit {\em normal forms}, which are obtained by showing that the languages of finite words involved in the classes' definitions can be chosen to be regular. $\LTL$-definable properties admit normal forms in $\LTL$ with past according to their belonging to the respective classes of the hierarchy too. The forms are given in \cite{MP89a}. Detailed proofs can be found in \cite{SafetyProgress10.1007/978-3-642-58041-3_5}. The proofs start from the assumption that a regular $\omega$-language $L$ is accepted by an appropriate Streett automaton. Then it is shown that, for each of the classes, a Streett automaton accepting $L$ with its accepting condition appropriately specialized exists. Finally, the relevant regular languages $M_1',M_1'',\ldots,M_N',M_N''$, are defined as the sets of words which take the specialized Streett automaton to a state from one of the sets that appear in its Streett acceptance condition.

\parag{Streett automata} (cf., e.g., \cite{DBLP:conf/dagstuhl/Farwer01}) are a type of $\omega$-automata. A deterministic Streett automaton with finite input alphabet $\Sigma$ has the form $\mathcal{A}\defeq \angles{Q,q_0,\delta,\{\angles{X_k,Y_k},k=1,\ldots,K\}}$ where $X_1,Y_1,\ldots,X_K,Y_K\subseteq Q$ and $\delta:Q\times\Sigma\rightarrow Q$. Deterministic Streett automata have the same expressive power as non-deterministic ones, which is not the case for, e.g., B\"uchi automata. Every regular $\omega$-language can be defined by a deterministic Streett automaton.
Given $\mathcal{A}$, let $\delta$ additionally denote its own extension to finite words as the second operand:
\[\delta(q,\angles{})\defeq q,\ \ \ \delta(q,\sigma^0\sigma^1\ldots)\defeq\delta(\delta(r,\sigma^0),\sigma^1\ldots),\ q\in Q.\]
Given $\sigma\in\Sigma^\omega$, let
\[\rho(\sigma)  \defeq \angles{\delta(q_0,\sigma^{0..n-1}):n<\omega}\]
Given and $\rho\in Q^\omega$, let
\[\Inf(\rho) \defeq  \{q\in Q:\rho^k=q\mbox{ for infinitely many }k\}.\]
$\mathcal{A}$ {\em accepts} $\sigma\in\Sigma^\omega$, if, for all $k\in\{1,\ldots,K\}$, 
$|Y_k\cap\Inf(\rho(\sigma))|<\omega$ implies $|X_k\cap\Inf(\rho(\sigma))|<\omega$.
In words, $\sigma$ is accepted, if, for all pairs $\angles{X_k,Y_k}$, $\sigma$ either makes $\mathcal{A}$ visit $Y_k$ infinitely many times, or makes $\mathcal{A}$ visit $X_k$ only finitely many times.

\parag{$\ITL$ with the neighbourhood modalities}
An in-depth introduction to propositional discrete-time $\ITL$, which has $\Next$, \emph{chop} and \emph{chop-star} as the temporal operators, can be found in \cite{Mos86,CMZ}. Given a vocabulary $V$ of {\em propositional variables}, $\ITL$ formulas that are written in $V$ are evaluated at finite or right-infinite sequences $\sigma=\sigma^0\sigma^1\ldots$ of {\em states} $\sigma^k\in\Sigma\defeq{\mathcal P}(V)$. The term {\em intervals} is used for such sequences in $\ITL$ and in its extensions. 

Along with $\Next$, \emph{chop} and \emph{chop-star}, $\ITLNL$ includes {\em neighbourhood modalities}, which are known as $\angles{\overline{A}}$ and $\angles{A}$ in Halpern-Shoham logic $\HS$ and its subsets such as propositional $\NL$. In this paper we avoid overloading $A$ by adopting the symbols $\Diamond_l$ and $\Diamond_r$ from the predicate real-time logics based on the neighbourhood modalities in \cite{DBLP:conf/compos/ChaochenH97,DBLP:journals/logcom/BaruaRC00} in $\ITLNL$:
\[
\mathsf{A} ::= \,\False \midd p \midd \mathsf{A} \Imp \mathsf{A} \midd \Next \mathsf{A} \midd \mathsf{A};\mathsf{A} \midd \mathsf{A}^*\midd \Diamond_l \mathsf{A}\midd \Diamond_r \mathsf{A}
\]
where $p$ stands for a propositional variable. We write $\Var(A)$ for the set of propositional variables occurring in formula $A$.

Satisfaction in the form $\sigma\models A$ as in $\ITL$ does not account for the possibility that time extends beyond the endpoints of $\sigma$ as necessary to define $\Diamond_l$ and $\Diamond_r$. Therefore, with $\sigma:\Zset\rightarrow\Sigma$, satisfaction has the form $\sigma,i,j\models A$. Here $i,j\in\Zset$ and $i\leq j$, define a {\em reference interval} within $\sigma$. This is equivalent to $\sigma^i\ldots\sigma^j\models A$ in the standard $\ITL$ notation for $A$ with no neighbourhood modalities. For the sake of simplicity we fix $\dom\sigma=\Zset$ and rule out infinite reference intervals, where $i=-\infty$, or $j=\infty$, in this paper. Generalizing the results to $\dom\sigma$ being any interval in $\Zset$ and allowing infinite reference intervals as in \cite{GM2022} can be achieved by straightforward modifications, which, however, would make the presentation more complicated. 

The defining clauses for $\models$ are as follows:
\[\begin{array}{lcl}
\sigma,i,j\not \models\False\\
\sigma,i,j\models p &\mbox{iff}& p\in\sigma^i\ \mbox{(by the locality principle)}\\
\sigma,i,j\models A\Imp B &\mbox{iff}& \sigma,i,j \models B\mbox{ or }\sigma,i,j \not\models A\\
\sigma,i,j\models \Next A &\mbox{iff}& i<j\mbox{ and } \sigma,i+1,j\models A\\
\sigma,i,j\models A;B &\mbox{iff}& \sigma,i,k\models A\mbox{ and }\sigma,k,j\models B \mbox{ for some }k\in\{i,\ldots,j\}\\
\sigma,i,j\models\Diamond_l A&\mbox{iff}& \mbox{ there exists a } k\leq i\mbox{ such that }\sigma,k,i\models A\\
\sigma,i,j\models\Diamond_r A&\mbox{iff}& \mbox{ there exists a } k\geq j\mbox{ such that }\sigma,j,k\models A
\end{array}\]
The definitions of $\models$ for $A^*$ (\emph{chop-star}) and propositional quantification $\exists p A$, whose interplay is key to establishing uniform Craig interpolation for $\ITLNL$ in this paper, are as follows. For $\sigma',\sigma'':\Zset\rightarrow\Sigma$, $\sigma'\sim_p\sigma''$, if $\sigma^k\setminus\{p\}=(\sigma')^k\setminus\{p\}$ for all $k\in\Zset$. Then
\[\begin{array}{lcl}
\sigma,i,j\models A^* &\mbox{iff}& i=j,\mbox{ or there exists a sequence }k_0=i<k_1<\ldots<k_N=j\\
& & \mbox{such that }\sigma,k_{n-1},k_n\models A\mbox{ for }n=1,\ldots,N.\\
\sigma,i,j\models\exists p A &\mbox{iff}& \mbox{there exists a }\sigma'\sim_p\sigma\mbox{ such that }\sigma',i,j\models A
\end{array}\]
The connectives of $\ITLNL$ include \emph{chop-star} but quantification, which is shown to be expressible below, is not included. 
Formula $A$ is {\em valid}, written $\models A$, if $\sigma,i,j\models A$ for all $\sigma$ and all pairs $i,j\in\Zset$, such that $i\leq j$. We write $\ITLNL$ for $\ITL$ with $\Diamond_l$ and $\Diamond_r$. 

The definitions of $\True$, $\neg$, $\wedge$, $\vee$ and $\equiv$ are as usual in $\ITLNL$.
We use the following defined temporal constants and derived operators below:
\[\begin{array}{lll}
\Empty \defeq \neg\Next\True&\mbox{- a single state interval}\\
\Skip \defeq \Next\Empty &\mbox{- a length 1 (2 states) interval}\\
\Prev A\defeq A;\Skip &A \mbox{ at the interval minus the last state}\\
\Diamond A\defeq \True;A,\qquad \Di A\defeq A;\True &A \mbox{ in some suffix subinterval}, A \mbox{ in some prefix subinterval}\\
\Box A\defeq\neg\Diamond\neg A,\qquad\Bi A\defeq\neg\Di\neg A &A \mbox{ in all suffix subintervals}, A \mbox{ in all prefix subintervals}\\
\Box_l A\defeq\neg\Diamond_l\neg A,\ \Box_r A\defeq\neg\Diamond_r\neg A&A \mbox{ in all adjacent intervals on the left, resp. on the right}\\
\Fin A\defeq\Diamond(A\wedge\Empty)& A \mbox{ in the single state suffix subinterval, i.e., in the last state}
\end{array}\]
Here follow the $\ITLNL$ connectives, basic and derived, listed in decreasing order of their binding strength:
\[\neg,\,\Next,\,\Prev,\,\Diamond_l,\,\Diamond_r,\,\Box_l,\,\Box_r,\Diamond,\Box,\mbox{ and }(.)^*;\ \ (.;.);\ \ \wedge;\ \ \vee;\ \ \Imp \mbox{ and }\equiv\ .\]
To access an arbitrary finite interval in a $\sigma:\Zset\rightarrow\Sigma$, one can use
\[\Diamond_a\defeq\Diamond_r\Diamond_r\Diamond_l\Diamond_l.\]
Its universal dual $\Box_a\defeq\neg\Diamond_a\neg$ is the {\em universal modality}  \cite{GorankoPassy1992} of $\ITLNL$.

In this paper we assume finite reference intervals, unless expressly stated otherwise. Right-infinite intervals are common in $\ITL$. With right-infinite intervals allowed, \emph{chop} admits both the strong form we use in $\ITLNL$:
\[\sigma\models A;B\mbox{ iff }\sigma^0\ldots\sigma^k\models A\mbox{ and }\sigma^k\ldots\models B \mbox{ for some }k\in\dom \sigma\ ,\]
and the weak form
\[\begin{array}{ll}
\sigma^0\sigma^1\ldots\models A;B& \mbox{iff either }\sigma^0\ldots\sigma^k\models A\mbox{ and }\sigma^k\ldots\models B \mbox{ for some }k\in\dom \sigma,\\
&\mbox{or }|\sigma|=\omega\mbox{ and }\sigma\models A\ .
\end{array}\]
This concurs with modeling sequential composition $A;B$ where $A$ may as well not terminate. The two variants of \emph{chop} are interdefinable, cf., e.g., \cite{GM2022}. In this paper we assume the strong variant everywhere. Finite intervals can be told apart from right-infinite ones using
\[\Finite\defeq\Diamond\Empty\ .\]
\parag{Expressiveness of $\ITL$ and $\ITLNL$}
The expressive completeness of introspective $\ITL$ on finite intervals was investigated in \cite{MosThesis}. With either \emph{chop-star} or propositional quantification, or both, introspective $\ITL$ is expressively complete wrt the monadic second order theory of finite sequences, or $\angles{\omega,<}$, if infinite intervals are allowed. With none of these connectives, expressive completeness holds wrt the respective monadic {\em first} order theories. The same holds about $\ITLNL$ wrt $\angles{\Zset,<}$ \cite{GM2024}, where infinite intervals are included. Other results on $\ITLNL$ such as separation apply to its $\exists$- and \emph{chop-star}-free subset too as shown in \cite{GM2022}. However, our proof of Craig interpolation as a corollary to the existence of strongest consequence formulas in this paper hinges on the expressibility of quantification with the use of \emph{chop-star}. We do not have this result for the \emph{chop-star}-free subset.

$\ITL$ can be viewed as the $\Diamond_l$- and $\Diamond_r$-free subset of $\ITLNL$ on finite or right-infinite intervals. Every regular language $L\subseteq\Sigma^+$ for $\Sigma$ being the powerset of some propositional vocabulary $V$, can be defined as the set of the finite intervals which satisfy some $\ITL$ formula written in $V$. The same applies to regular $\omega$-languages \cite{MosThesis}.

\parag{Optional Negations}
In the sequel we often use $\varepsilon$ to denote an {\em optional negation}, that is, either $\neg$, or nothing. We write $\overline{\varepsilon}$ for the alternative optional negation of $\varepsilon$.

\parag{Guarded Normal Form (GNF)} Temporal logics and process algebras commonly admit {\em guarded normal forms} which are case distinctions featuring a full system of first state/initial process step options and the respective possible continuations. In both $\LTL$ and $\ITL$ the form of formula $A$ reads
\begin{equation}\label{gnf}
A_e\wedge\Empty\vee\bigvee\limits_{k=1}^K A_k\wedge\Next A_k'\ ,
\end{equation}
the key difference being that temporal connectives which, unlike $\Next$, are not shared by $\LTL$ and $\ITL$, can appear in the $A_k'$s. $A_e$ and the $A_k$s are supposed to be {\em state}, i.e., modality-free. $A_k$ are supposed to be pairwise inconsistent, and can be guaranteed to be a full system, which we assume below, possibly at the cost of adding a disjunct
of the form $\neg\bigl(\bigvee\limits_{k=1}^K A_k\bigr)\wedge\Next\False$. The finest (and costliest to process) GNF commences, if every conjunction of the form $\bigwedge\limits_{p\in\Var(A)}\varepsilon_p p$ appears as an $A_k$. The coarsest GNF satisfies the additional condition that no two $A_k'$s are equivalent and can be obtained from any other GNF by merging the disjuncts which have equivalent $A_k'$s. Both the finest and the coarsest GNF are unique up to equivalence of the $A_k$s and the $A_k'$s. Since the $A_k$s form a full system, the equivalence of (\ref{gnf}) to $A$ implies
\begin{equation}\label{gnfUniversal}
\models A\equiv( \Empty\Imp A_e)\wedge\bigwedge\limits_{k=1}^K (A_k\wedge\neg\Empty)\Imp\Next A_k'
\end{equation}
For \emph{chop-star}-free $A$, $A_k'$ can be chosen to be \emph{chop-star}-free too.
GNF was introduced to $\ITL$ in \cite{DBLP:phd/ethos/Duan96}.

\parag{Reversing (aka mirroring) time}
Observe that $\Prev$ is the time inverse of $\Next$ because:
\[\sigma,i,j\models A;\Skip\mbox{ iff }i<j\mbox{ and }\sigma,i,j-1\models A.\]
This kind of symmetry lets us avoid repeats by referring to the {\em time inverses} $A^{-1}$ of $\ITLNL$ formulas $A$ after Section 6 from \cite{DBLP:conf/csl/Rasmussen99}. By definition, $\ldots \sigma^{k-1}\sigma^k\sigma^{k+1}\ldots,i,j\models A^{-1}$ iff $\ldots \sigma^{k+1}\sigma^k\sigma^{k-1}\ldots,j,i\models A$. Hence $\models A$ and $\models A^{-1}$ are equivalent. In $\ITLNL$, $A^{-1}$ can be expressed using the validity of the equivalences:
\[\begin{array}{lll}
\False^{-1} \equiv \False & p^{-1} \equiv \Fin p& (A\Imp B)^{-1} \equiv A^{-1}\Imp B^{-1}\\
(\Next A)^{-1}  \equiv \Prev(A^{-1}) & (A;B)^{-1} \equiv B^{-1};A^{-1} & (A^*)^{-1} \equiv (A^{-1})^*\\
(\Diamond_l A)^{-1}\equiv \Diamond_r A^{-1} & (\Diamond_r A)^{-1} \equiv \Diamond_l A^{-1}
\end{array}\]

\parag{The separation theorem for $\ITLNL$ \cite{GM2022}} Formulas with no occurrences of $\Diamond_l$ or $\Diamond_r$ are called {\em introspective} in $\ITLNL$. Along with introspective formulas, separation refers to past and future $\ITLNL$ formulas.
\begin{definition}
\label{futurepastformulas}
An $\ITLNL$ formula $F$ is (non-strictly) {\em future} if it has the syntax
\[\mathsf{F} ::= \mathsf{C}\midd \neg \mathsf{F} \midd \mathsf{F} \vee \mathsf{F} \midd \Diamond_r \mathsf{F}\]
where $\mathsf{C}$ denotes the class of introspective, that is, $\ITL$ formulas. The syntax of non-strictly {\em past} formulas is similar, with $\Diamond_r$ replaced by $\Diamond_l$. 
\end{definition}
In words, future formulas are $\Diamond_l$-free $\ITLNL$ formulas where $\Diamond_r$ is not allowed in the scope of \emph{chop} or \emph{chop-star}.

{\em Strictly} future and {\em strictly} past formulas are additionally restricted to impose no conditions on the reference interval. The designated occurrences of $\Skip$ in them are meant to avoid the end-points of the reference interval:
\begin{definition}
An $\ITLNL$ formula is {\em strictly future (past)}, if it has the form $\Diamond_r(\Skip;F)$ ($\Diamond_l(P;\Skip)$) where $F$ is future ($P$ is past).
\end{definition}
Syntactically, {\em strictly} future (past) formulas are not future (past) in the sense of Definition (\ref{futurepastformulas}). However, every {\em strictly} future (past) formula is equivalent to a future (past) one \cite{GM2022}.

A {\em strictly separated formula} is a Boolean combination of strictly past, introspective and strictly future formulas.\begin{theorem}[Corollary 2.6 in \cite{GM2022}]\label{septhmmaincor}
Let $A$ be an $\ITLNL$ formula. Then there exists a strictly separated $S$ such that $\models A\equiv S$.
\end{theorem}
Importantly, the proof of Theorem \ref{septhmmaincor} in \cite{GM2022} is based on syntactical transformations by means of valid equivalences. Hence the equivalence of any given formula and its separated form can be easily made part of a proof in any complete proof system for $\ITLNL$. These transformations are valid in this paper's variant of $\ITLNL$ where infinite reference intervals are excluded too. Therefore the theorem applies without change.

Along with Theorem \ref{septhmmaincor}, in this paper we use one more proposition from \cite{GM2022}, which can be described as an {\em introspective} form of separation as it is about writing an $\ITL$, that is, an $\Diamond_l$- and $\Diamond_r$-free formula as a combination of conditions on the possible choppings of reference intervals into prefix and suffix parts, with these conditions written in $\ITL$ too. $\ITL$ formulas $A_1,\ldots,A_K$ form a {\em full system}, if $\models\bigvee\limits_{k=1}^K A_k$, and $\models\neg(A_{k_1}\wedge A_{k_2})$ for $k_1\not=k_2$.

\begin{proposition}[Lemma 3.2 in \cite{GM2022}]\label{keyprop}
Let $A$ be an $\ITL$ formula. Then there exists a $K< \omega$ and some $\ITL$ formulas $A_k,A_k'$, $k=1,\ldots,K$, such that $A_1,\ldots,A_K$ is a full system and
\begin{equation}\label{keyEq}
\models A\equiv\bigvee\limits_{k=1}^K A_k; A_k'\quad\models A\equiv\bigwedge\limits_{k=1}^K\neg(A_k;\neg A_k').
\end{equation}
\end{proposition}
The time-inverse statement holds too:
\begin{proposition}[Lemma 3.4 in \cite{GM2022}]\label{keypropMirror}
Let $A$ be an $\ITL$ formula. Then there exists an $K<\omega$ and some $\ITL$ formulas $A_k,A_k'$, $k=1,\ldots,K$, such that $A_1,\ldots,A_K$ is a full system and
\[\models A\equiv\bigvee\limits_{k=1}^K A_k';A_k\quad\models A\equiv\bigwedge\limits_{k=1}^K\neg(\neg A_k';A_k).\]
\end{proposition}
In \cite{GM2022}, these two propositions are part of the proof of Theorem \ref{septhmmaincor}, which covers expanding modalities. Importantly, the equivalences (\ref{keyEq}) from Proposition \ref{keyprop} admit the following {\em strict} form:
\begin{equation}\label{keyEqStrict}
\models A\equiv A_e\wedge\Empty\vee\bigvee\limits_{k=1}^K A_k;\Skip;A_k'\quad\models A\equiv(\Empty\Imp A_e)\wedge\bigwedge\limits_{k=1}^K\neg(A_k;\Skip;\neg A_k').
\end{equation}

\parag{The Projection Operator $\Pi$ in $\ITL$ \cite{DBLP:conf/icalp/HalpernMM83}}
Given a $\sigma\in\Sigma^+$, $\Sigma\defeq\mathcal{P}(V)$, and $X\subseteq\Sigma$, we write $\sigma|_X$ for the sequence obtained by deleting the states $\sigma^i\not\in X$ from $\sigma$. For state formulas $w$, $\sigma|_w\defeq \sigma|_{\{x\in\Sigma:\ x\models w\}}$. Given a state formula $w$ and an $\ITL$ formula $A$,
\[\sigma\models w\Pi A\mbox{ if }\sigma|_w\models A.\]
This tacitly implies that $\sigma\not\models w\Pi A$ for $\sigma$ with no $w$-states in them because $\sigma|_w$ is the empty sequence for such $\sigma$.
With its left operand $w$ fixed, $\Pi$ is a $\mathbf{D_c}$-modality (cf. e.g. \cite{hughes1996new}) because $\sigma\mapsto \sigma|_w$ is a partial function. It admits a {\em weak} dual $w\overline{\Pi}A\defeq\neg(s\Pi(\neg A))$, which is trivially true, if $\sigma|_w$ is the empty sequence.  

$\Pi$ can be defined for {\em temporal} formulas $W$ as the left operand too. In that case $\sigma|_W$ is the result of deleting $\sigma^i$ for $i\leq |\sigma|$ such that $\sigma^i\sigma^{i+1}\ldots\not\models\ W$. Then $\sigma\models W\Pi A$ holds iff $\sigma|_W\models A$ again. There is no single natural way to extend the definition of $W\Pi A$ with temporal $W$ to satisfaction of the form $\sigma,i,j\models\ldots$ as in $\ITLNL$. The case of $w$ being a state formula is sufficient for our purposes in this paper.

\section{A Normal Form for Reactivity in $\ITLNL$}
\label{reactivityNF}

This is one of the most straightforward applications of Proposition \ref{keyprop}. It is known that given a regular $\omega$-language $L\subseteq\Sigma^\omega$, $\Sigma\defeq\mathcal{P}(V)$, there exists an $\ITL$ formula $F\defeq F_L$ in the vocabulary $V$ such that $L=\{\sigma\in\Sigma^\omega:\sigma\models F\}$, and vice versa. The same applies to languages of non-empty finite words and $\ITL$ on finite intervals. Given the regular language definition (\ref{reactReg}) for reactivity from \cite{MP89a}, we can assume that there exist some $\ITL$ formulas $F_{L,1}',\ldots,F_{L,N}'$ and $F_{L,1}'',\ldots,F_{L,N}''$ such that
\[
\begin{array}{ll}
\sigma\in L\mbox{ iff} &\mbox{for some }n\in\{1,\ldots,N\}\\
& \mbox{there exist only finitely many $l$ such that }\sigma^{0..l}\models\neg F_{l,n}'\\
& \mbox{and infinitely many $l$ such that }\sigma^{0..l}\models F_{l,n}''.
\end{array}\] 
Let $C$ be any $\ITL$ formula, and let $\bigvee\limits_{k=1}^K C_k;C_k'$ meet the requirements of Proposition \ref{keyprop}. Then
\begin{proposition}\label{proReact}
The number of the prefixes of a $\sigma:\omega\rightarrow\Sigma$ which satisfy $C$ is finite iff any extension $\sigma':\Zset\rightarrow \Sigma$ of $\sigma$ satisfies 
$\sigma',0,0\models\Diamond_r\bigvee\limits_{k=1}^K(C_k\wedge\Box_r\neg C_k')$.
\end{proposition}
In Proposition \ref{proReact}, we need $\sigma'$ just to conform with the definition of $\models$ in $\ITLNL$; no reference to the states in $\sigma'$ on the left of $0$ is to be made. 

\begin{proof}
Observe that, if $\sigma^{0..l}\models C_k$, then, for any $l'\geq l$ to satisfy $\sigma^{0..l'}\models C$, it is necessary that $\sigma^{l..l'}\models C_k'$, and this is denied by $\sigma',0,l\models\models\Box_r\neg C_k'$. Hence $\sigma',0,0\models\Diamond_r(C_k\wedge\Box_r\neg C_k')$ implies that an $l<\omega$ exists such that $\sigma^{0..l'}\not\models C$ for all $l'\geq l$. Conversely, let $l<\omega$ be such that $\sigma^{0..l'}\not\models C$ for all $l'\geq l$. Since $C_1,\ldots,C_K$ form a full system, there exists a $k\in\{1,\ldots,K\}$ such that $\sigma^{0..l}\models C_k$, and assuming that $\sigma',0,l\models\Diamond C_k'$ is a contradiction because implies $\sigma,0,l'\models C_k;C_k'$, and $\models C_k;C_k'\Imp C$. Hence there exists  a $k\in\{1,\ldots,K\}$ such that $\sigma',0,0\models\Diamond_r(C_k\wedge\Box_r\neg C_k')$.
\end{proof}

\qquad

\noindent
Given a regular language $X$ of nonempty words, we let 
\[\mathit{Fin}(X)\defeq \Diamond_r\bigvee\limits_{k=1}^K(C_{X,k}\wedge\Box_r\neg C_{X,k}')\]
where $C_X$ is some arbitrary fixed $\ITL$ formula which defines $X$ and $C_{X,k}$ and $C_{X,k}'$ are as above.
Using this notation, (\ref{reactReg}) translates into the following defining equivalence for the considered regular $\omega$-language $L$ by an $\ITLNL$ formula:
\[\sigma\in L\mbox{ iff }\sigma',0,0\models\bigvee\limits_{n=1}^N\mathit{Fin}(\Sigma^*\setminus M_n')\wedge\neg\mathit{Fin}(M_n'')\]
where $\sigma'$ is an arbitrary extension of $\sigma$ `into the past' as in Proposition \ref{proReact}. Similarly (\ref{reactRegConj}) translates into
\begin{equation}\label{ITLNLreactNFConj}
\sigma\in L\mbox{ iff }\sigma',0,0\models\bigwedge\limits_{n=1}^N\mathit{Fin}( M_n'')\Imp\mathit{Fin}(M_n').
\end{equation}

\section{Separation with Additional Conditions and a Normal Form for Finite-time $\ITL$}

In this section we show how a special form of (\ref{keyEqStrict}) can be used to obtain a normal form for $\ITL$. Given an $\ITL$ formula $A$ and a state formula $w$, the normal form of $A$ allows reading the conditions to be satisfied by the maximal $\Box w$- and $\Box\neg w$-subintervals of any $\sigma\in\Sigma^+$ such that $\sigma\models A$. As it becomes clear below, the normal form is built starting from formulas of the forms $B\wedge \Box w$ and $B\wedge\Box\neg w$ using $\vee$, \emph{chop} and \emph{chop-star}. 

Consider (\ref{keyEqStrict}) with a focus on $(A_k;\Skip; A_k')$. Let, given some state formula $w$, $A_k$ be additionally restricted to be satisfied in a $\Box w$-interval. Let $A_k'$ be restricted to be satisfied at an interval where $\neg w$ holds initially. We claim that there exist some $\ITL$ formulas and $A_k^w,\ A_k^{\neg w,'}$, $k=1,\ldots,K$, such that
$A_1^w,\ldots, A_K^w$ form a full system and
\begin{equation}\label{keyEqStrictH}
\begin{array}{l}
\models A\wedge w\equiv A\wedge\Box w\vee
\ \bigvee\limits_{k=1}^K (A_k^w\wedge\Box w);\Skip;(A_k^{\neg w,'}\wedge\neg w)\\
\qquad\\
\models A\wedge w\equiv (\Box w\Imp A)\wedge\bigwedge\limits_{k=1}^K\neg\bigg((A_k^w\wedge\Box w);\Skip;\neg(A_k^{\neg w,'}\wedge\neg w)\bigg).
\end{array}
\end{equation}
The formulas $A_1^w,\ldots,A_K^w$ in (\ref{keyEqStrictH}) are supposed to be a full system {\em relative to $\Box w$}. This means that $\vdash\Box w\Imp\bigvee\limits_{k=1}^K A_k^w$ and $\vdash\neg(A_{k_1}^w\wedge A_{k_2}^w\wedge\Box w)$ for $k_1\not=k_2$. The equivalences (\ref{keyEqStrictH}) can be written with $\neg w$ in the role of $w$ too. Below we prove that, given $A$ and $w$, some finite sets of $\ITL$ formulas $\Cl^w(A)$ and $\Cl^{\neg w}(A)$ can be found such that $A\wedge\varepsilon w\in \Cl^{\varepsilon w}(A)$, and every formula $B\in\Cl^{\varepsilon w}(A)$ can be written as in (\ref{keyEqStrictH}) using some $B_k^{\overline{\varepsilon}w,'}\wedge\overline{\varepsilon} w\in\Cl^{\overline{\varepsilon} w}(A)$. Finally we use all the equivalences of the form (\ref{keyEqStrictH}) for formulas $B\in \Cl^w(A)\cup Cl^{\neg w}(A)$ as a system of canonical equations. We obtain the desired normal form of $A$ by solving this system. We do the details for the case of right-infinite intervals. The case of finite intervals is simpler and can be done similarly. 
\begin{proposition}\label{keyPropH2}
For any $\ITL$ formula $A$ there exists a pair of finite sets $\mathcal{F}^w$ and $\mathcal{F}^{\neg w}$ such that $A\wedge w\in\mathcal{F}^w$, $A\wedge\neg w\in \mathcal{F}^{\neg w}$ and, for any optional negation $\varepsilon$, if $B\in\mathcal{F}^{\varepsilon w}$, then
\begin{equation}\label{keyEqStrictH2}
\begin{array}{l}
\models B\wedge \varepsilon w\equiv B\wedge\Box\varepsilon w\vee
\ \bigvee\limits_{k=1}^K (B_k^{\varepsilon w}\wedge\Box\varepsilon w);\Skip;(B_k^{\overline{\varepsilon} w,'}\wedge\overline{\varepsilon} w)\\
\qquad\\
\models B\wedge \varepsilon w\equiv (\Box\varepsilon w\Imp B)\wedge\bigwedge\limits_{k=1}^K\neg\bigg((B_k^{\varepsilon w}\wedge\Box\varepsilon w);\Skip;\neg(B_k^{\overline{\varepsilon} w,'}\wedge\overline{\varepsilon} w)\bigg).
\end{array}
\end{equation}
for some $B_1^{\varepsilon w},\ldots,B_K^{\varepsilon w}$ which form a full system relative to $\Box\varepsilon w$, and some $B_1^{\overline{\varepsilon} w,'},\ldots,B_K^{\overline{\varepsilon} w,'}\in\mathcal{F}^{\overline{\varepsilon} w}$. 
\end{proposition}

\begin{proof}
Given Streett automaton $\mathcal{A}\defeq\angles{Q,q_0,\delta,\{\angles{X_n,Y_n}:n=1,\ldots,N\}}$ which accepts the language defined by $A$, let
\[\mathcal{A}_{w,\neg w}\defeq \angles{Q_{w,\neg w},{\angles{q_0,*}},\delta_{w,\neg w},\mathit{Acc}_{w,\neg w}}\]
where 
\[\begin{array}{llll}
\multicolumn{4}{l}{Q_{w,\neg w}\defeq Q\times\{*,w,\neg w,1\}};\\
\delta_{w,\neg w}(\angles{q,*},x)&\defeq&\angles{\delta(q,x),\varepsilon w}, &\mbox{if }x\models \varepsilon w;\\
\delta_{w,\neg w}(\angles{q,\varepsilon w},x)&\defeq&\angles{\delta(q,x),\varepsilon w}, &\mbox{if }x\models\varepsilon w;\\
\delta_{w,\neg w}(\angles{q,\varepsilon w},x)&\defeq&\angles{\delta(q,x),1}, &\mbox{if }x\models\overline{\varepsilon} w;\\
\delta_{w,\neg w}(\angles{q,1},x)&\defeq&\angles{\delta(q,x),1};\\
\multicolumn{4}{l}{\mathit{Acc}_{w,\neg w}\defeq\{\angles{X_n\times\{*,w,\neg w,1\},Y_n\times\{*,w,\neg w,1\}}:n=1,\ldots,N\}.}
\end{array}\]
Along with $\mathcal{A}_{w,\neg w}$, below we consider the automata $\mathcal{A}_{w,\neg w,\angles{q,*}}\defeq\angles{Q_{w,\neg w},{\angles{q,*}},\delta_{w,\neg w},\mathit{Acc}_{w,\neg w}}$ obtained by replacing the initial state $\angles{q_0,*}$ of $\mathcal{A}_{w,\neg w}$ with $\angles{q,*}$, $q\in Q$.
In these automata, states of the form $\angles{q,*}$ play the role of initial states, states of the form $\angles{q,\varepsilon w}$ are reached by input words which satisfy $\varepsilon w$ all along, and states of the form $\angles{q,1}$ are reached as soon as a switch between $w$- and $\neg w$-input letters occurs in the input word.

Let
\[L_{\angles{q',*},\angles{q'',\varepsilon w}}\defeq\{\sigma\in\Sigma^+:\delta_{w,\neg w}(\angles{q',*},\sigma)=\angles{q'',\varepsilon w}\}, \ q',q''\in Q,\]
and let $L_{\angles{q,*}}$ be the language accepted by  $\mathcal{A}_{w,\neg w,\angles{q,*}}$, $q\in Q$. (A direct check shows that $L_{\angles{q,*}}=L_{\angles{q,w}}=L_{\angles{q,\neg w}}=L_{\angles{q,1}}$.)
$\mathcal{A}$ and $\mathcal{A}_{w,\neg w}=\mathcal{A}_{w,\neg w,\angles{q_0,*}}$ accept the same language $L_{\angles{q_0,*}}$, which is the language defined by $A$ because the states $\angles{q,*}$, $\angles{q,w}$, $\angles{q,\neg w}$ and $\angles{q,1}$ always appear together in the sets of states in $\mathit{Acc}_{w,\neg w}$. Let
\[L_r^{\varepsilon w,'}\defeq\bigcup\limits_{x\in\Sigma,\ \delta_{w,\neg\omega}(\angles{r,\overline{\varepsilon} w},x)=\angles{s,1}} x\cdot L_{\angles{s,1}},\ r\in Q.\]
The condition $\delta_{w,\neg\omega}(\angles{r,\overline{\varepsilon} w},x)=\angles{s,1}$ in the definition of $L_r^{\varepsilon w,'}$ implies that a summand for $x\in\Sigma$ appears in the defining expression iff $x\models\varepsilon w$. Let $B_q$ be an $\ITL$ formula which defines the $\omega$-language accepted by $\mathcal{A}_{\angles{q,*}}$. Let $B_q^{\varepsilon w,'}$ be an $\ITL$ formula which defines the $\omega$-language $L_q^{\varepsilon w,'}$, $q\in Q$.  Let $B_{q',q''}^{\varepsilon w}$ be an $\ITL$ formula which defines the finite word language $L_{\angles{q',*},\angles{q'',\varepsilon w}}$. A direct check shows that
\begin{equation}\label{binClosure0}
L_{\angles{q,*}}=L_q^{w,'}\uplus L_q^{\neg w,'},
\end{equation}
and
\begin{equation}\label{binClosure}
L_q^{\varepsilon w,'}\,=\,\{\sigma\in\Sigma^\omega:\sigma\models B_q\wedge\varepsilon w\}\,=\,
\{\sigma\in\Sigma^\omega:\sigma\models B_q\wedge\Box\varepsilon w\}
\,\cup\,\bigcup\limits_{r\in Q}L_{\angles{q,*},\angles{r,\varepsilon w}}\cdot {L_r^{\overline{\varepsilon} w,'}}
\end{equation}
In terms of the $\ITL$ formulas $B_q$, $B_q^{\varepsilon w,'}$ and $B_{q',q''}^{\varepsilon w}$, (\ref{binClosure0}) and (\ref{binClosure}) imply
\[\models B_q\equiv B_q^{w,'}\vee B_q^{\neg w,'}\mbox{ and }
\models B_q^{\varepsilon w,'}\wedge\varepsilon w\equiv 
B_q^{\varepsilon w,'}\wedge\Box \varepsilon w\vee
\bigvee\limits_{q'\in Q}
(B_{q,q'}^{\varepsilon w}\wedge\Box \varepsilon w);\Skip;(B_q^{\overline{\varepsilon} w,'}\wedge\overline{\varepsilon}w),\]
respectively. We define $\mathcal{F}^{\varepsilon w}\defeq \{B_q^{\varepsilon w,'}:q\in Q\}$.
\end{proof}

\qquad

\noindent
In the sequel, given $A$, we assume $\Cl^w(A)$ and $\Cl^{\neg w}(A)$ to denote some fixed pair of finite sets $\mathcal{F}^w$ and $\mathcal{F}^{\neg w}$ that satisfy the conditions of Proposition \ref{keyPropH2}.

\begin{remark}\label{remSat}
Importantly, all the formulas $B_k^{\varepsilon w}$ in (\ref{keyEqStrictH2}) can be assumed to be satisfiable.
\end{remark}

Given and $\ITL$ formula $A$ and a state formula $w$, the equivalences (\ref{keyEqStrictH2}) for $B\in \Cl^{w}(A)\cup\Cl^{\neg w}(A)$ can be used as canonical equations, as known from the theory of formal languages, to derive a formula which is equivalent to $A$ and has the syntax
\begin{equation}\label{nfSyntax}
w\wedge(\mathsf{R}^{w,w}\vee R^{w,\neg w})\vee \neg w\wedge(\mathsf{R}^{\neg w,w}\vee R^{\neg w,\neg w})
\end{equation}
where:
\begin{equation}\label{regPosPM}
\begin{array}{l}
\mathsf{L}^{\varepsilon w,\varepsilon w}\ ::= (C\wedge\Box\varepsilon w);\Skip\midd \mathsf{L}^{\varepsilon w,\varepsilon w}\vee \mathsf{L}^{\varepsilon w,\varepsilon w}\midd \mathsf{L}^{\varepsilon w,\varepsilon' w};\mathsf{L}^{\overline{\varepsilon'}w,\varepsilon w}\\
[2mm]
\mathsf{L}^{\varepsilon w,\overline{\varepsilon} w}\ ::= \mathsf{L}^{\varepsilon w,\overline{\varepsilon} w}\vee \mathsf{L}^{\varepsilon w,\overline{\varepsilon} w}\midd \mathsf{L}^{\varepsilon w,\varepsilon' w};\mathsf{L}^{\overline{\varepsilon'} w,\overline{\varepsilon} w}\midd (\mathsf{L}^{\varepsilon w,\overline{\varepsilon} w})^*\\
[2mm]
\mathsf{R}^{\varepsilon w,\varepsilon w}\ ::= B\wedge\Box\varepsilon w\midd \mathsf{R}^{\varepsilon w,\varepsilon w}\vee \mathsf{R}^{\varepsilon w,\varepsilon w}\midd \mathsf{L}^{\varepsilon w,\varepsilon' w};\mathsf{R}^{\overline{\varepsilon'}w,\varepsilon w}\\
[2mm]
\mathsf{R}^{\varepsilon w,\overline{\varepsilon} w}\ ::= \mathsf{R}^{\varepsilon w,\overline{\varepsilon} w}\vee \mathsf{R}^{\varepsilon w,\overline{\varepsilon} w}\midd \mathsf{L}^{\varepsilon w,\varepsilon' w};\mathsf{R}^{\overline{\varepsilon'} w,\overline{\varepsilon} w}
\end{array}
\end{equation}
In the BNF for $\mathsf{R}^{\varepsilon w,\varepsilon w}$, $B$ ranges over $\Cl^{\varepsilon w}(A)$. In the BNF for $\mathsf{R}^{\varepsilon w,\varepsilon w}$, $C$ is one of the formulas $B_k^{\varepsilon w}$ appearing in the equivalences about the formulas from $\Cl^{\varepsilon w}(A)$ in (\ref{keyEqStrictH2}). The formulas $B\wedge\Box\varepsilon w$ and $C\wedge\Box\varepsilon w$ require the reference interval to satisfy $\varepsilon w$ all along. Formulas of the syntax $\mathsf{L}^{\varepsilon' w,\varepsilon'' w}$ or $\mathsf{R}^{\varepsilon' w,\varepsilon'' w}$ require the reference interval to have a prefix and a  suffix which satisfy $\Box\varepsilon' w$ and $\Box\varepsilon'' w$, respectively. Hence use of \emph{chop} in (\ref{regPosPM}) is restricted to formulas which require a $\Box\varepsilon w$ suffix on the left and formulas which require a $\Box\overline{\varepsilon}w$ prefix on the right. The designated $\Skip$ in $(C\wedge\Box\varepsilon w);\Skip$ guarantees that these subintervals do share end states. Indeed, it can easily be shown that all formulas of the syntax $\mathsf{L}^{\varepsilon w,\varepsilon' w}$ are equivalent to formulas of one of the forms $F;\Skip$ and $\Empty\vee F;\Skip$. The alternation between $w$-states and $\neg w$ states on the two sides of $;$ guarantees that, given $\sigma\models A$ where $A$ has the syntax $\mathsf{R}^{\varepsilon' w,\varepsilon'' w}$, $A$'s subformulas of the syntax $B\wedge\Box\varepsilon w$ specify properties of the subintervals of $\sigma$ that are {\em maximal} wrt the condition of satisfying $\varepsilon w$ all along.

Solving the system (\ref{keyEqStrictH2}) gives an equivalent of the given formula $A$ of the syntax (\ref{nfSyntax}). We do the details to let the reader check, by following the transformations below, that the solutions for all the unknowns $B\wedge \varepsilon w$, $B\in\Cl^{\varepsilon w}(A)$ really have the syntax $\varepsilon w\wedge(\mathsf{R}^{\varepsilon w, w}\vee \mathsf{R}^{\varepsilon w,\neg  w})$. 

The general form of the equations obtained at the various stages of solving (\ref{keyEqStrictH2}) with the syntax of the designated formulas specified is
\begin{equation}\label{intermediateCanonical}
B\wedge \varepsilon w\equiv \underbrace{R}_{\mathsf{R}^{\varepsilon w, w}\vee\mathsf{R}^{\varepsilon w,\neg w}}
\vee \bigvee\limits_{k=1}^K\  \underbrace{L_k}_{\mathsf{L}^{\varepsilon w,\overline{\varepsilon_k} w}};(B_k\wedge\varepsilon_k w)\ .
\end{equation}
Here $\varepsilon$ and $\varepsilon_k$, $k=1,\ldots,K$, are possibly different optional negations and $B_k\in\Cl^{\varepsilon_k w}(A)$. 

Solving such systems of equations is known as Gauss elimination in the theory of formal languages. At every step of solving (\ref{keyEqStrictH2}), an unknown $B\wedge \varepsilon w$ becomes eliminated using the equation (\ref{intermediateCanonical}) where that unknown appears on the LHS of $\equiv$. By letting $L_k\defeq\False$ for some $k$, we can assume that $B\wedge\varepsilon w$ appears on both sides of (\ref{intermediateCanonical}). Let (\ref{intermediateCanonical}) be
\[B\wedge \varepsilon w\equiv  \underbrace{R}_{\mathsf{R}^{\varepsilon w, w}\vee\mathsf{R}^{\varepsilon w,\neg w}}\vee \underbrace{L_1}_{\mathsf{L}^{\varepsilon w,\overline{\varepsilon} w}};(B\wedge \varepsilon w)\vee \bigvee\limits_{k=2}^K \underbrace{L_k}_{\mathsf{L}^{\varepsilon w,\overline{\varepsilon_k} w}};(B_l\wedge\varepsilon_l w)\ ,\]
that is, let $B\wedge \varepsilon w$ be $B_1\wedge \varepsilon_1 w$. Then $B\wedge\varepsilon w$ can be expressed using the equivalence
\[B\wedge \varepsilon w\equiv \underbrace{L_1^*}_{\mathsf{L}^{\varepsilon w,\overline{\varepsilon} w}};\left(\underbrace{R}_{\mathsf{R}^{\varepsilon w, w}\vee\mathsf{R}^{\varepsilon w,\neg w}}\vee \bigvee\limits_{k=2}^K\  \underbrace{L_k}_{\mathsf{L}^{\varepsilon w,\overline{\varepsilon_k} w}};(B_k\wedge\varepsilon_k w)\right)\ .\]
Since $R_1$ has the syntax $\mathsf{R}^{\varepsilon w,\overline{\varepsilon} w}$ and the syntax of $R$ and $R_l$, $l=2,\ldots,L$, is as above, the expression we obtain for $B\wedge \varepsilon w$ has the syntax $\mathsf{R}^{\varepsilon w, w}\vee\mathsf{R}^{\varepsilon w,\neg w}$ as required. Using this expression to eliminate $B\wedge\varepsilon w$ from instances
\[B'\wedge \varepsilon' w\equiv \underbrace{R'}_{\mathsf{R}^{\varepsilon' w, w}\vee\mathsf{R}^{\varepsilon' w,\neg w}}\vee \underbrace{L_1'}_{\mathsf{L}^{\varepsilon' w,\overline{\varepsilon} w}};(B\wedge \varepsilon w)\vee \bigvee\limits_{k=2}^K \underbrace{L_k'}_{\mathsf{L}^{\varepsilon' w,\overline{\varepsilon_k'} w}};(B_k'\wedge\varepsilon_k' w)\ .\]
of (\ref{intermediateCanonical}) for other $B'$ and $\varepsilon'$ produces
\[B'\wedge \varepsilon' w\equiv R'\vee L_1';\left(L_1^*;\left(R\vee \bigvee\limits_{k=2}^K L_k;(B_k\wedge\varepsilon_k w)\right)\right)\vee \bigvee\limits_{k=2}^K R_k';(B_k'\wedge\varepsilon_k' w)\ ,\]
which can be written as 
\[B'\wedge \varepsilon' w\equiv 
\underbrace{R'\vee 
L_1';L_1^*;R}_{\mathsf{R}^{\varepsilon' w, w}\vee\mathsf{R}^{\varepsilon' w,\neg w}}  \vee
\bigvee\limits_{k=2}^K\underbrace{((L_1';L_1^*;L_k)\vee L_k')}_{\mathsf{L}^{\varepsilon' w,\overline{\varepsilon_k'} w}}
;(B_k'\wedge\varepsilon_k' w)\ .\]
We have proved the following theorem:
\begin{theorem}\label{itlNFTheorem}
Every $\ITL$ formula $A$ is equivalent to a formula of the syntax (\ref{nfSyntax}).
\end{theorem}

\section{The Inverse of $\Pi$}

Let the {\em inverse of $\Pi$ wrt its second operand} be defined as follows:
\[\sigma\models w\PiInv A,\mbox{ if there exists a }\sigma'\mbox{ such that }\sigma'|_w=\sigma\mbox{ and }\sigma'\models A\]
where $\sigma'|_w$ is as in the definition of $\Pi$ in the Preliminaries section. In words, $\sigma\models w\PiInv A$, if $\sigma$ can be obtained by erasing the non-$w$ states from some sequence $\sigma'$ which satisfies $A$.
It can easily be shown that
\[\models w\PiInv(w\Pi A)\equiv A\wedge\Box w\mbox{ and }\models A\wedge\Box w\Imp w\Pi(w\PiInv A)\ .\]
$w\Pi(w\PiInv A)\Imp A$ is not valid because $w\Pi(w\PiInv A)$ is true in all intervals whose $w$-states form an interval that satisfies $w\PiInv A$, whereas $A$ can impose conditions on the $\neg w$-parts of such intervals as well. 
\begin{theorem}\label{theoremInvPi}
Given an $\ITL$ formula $A$ and a state formula $w$, a $\PiInv$-free $\ITL$ formula $A'$ that is equivalent to $w\PiInv A$.
\end{theorem}
The proof of Theorem \ref{theoremInvPi} refers to the validity of several equivalences about $\PiInv$.
\begin{lemma}\label{lemmaInvPi}
 Let $w$ denote a state formula. Then
\begin{eqnarray}
\label{piInverseProps1}
\models w\PiInv(B\wedge\Box w)\equiv B\wedge\Box w\qquad\\
\label{piInverseProps2}
\models( w\PiInv(B\wedge\Box\neg w))\equiv\False\qquad\\
\label{piInverseProps3}
\models w\PiInv((B\wedge\Box\neg w);\Skip;(A\wedge w))\equiv 
\left\{\begin{array}{ll}
\False,\mbox{ if }\models\Finite\Imp\neg (B\wedge\Box\neg w);\\
w\PiInv(A\wedge w),\mbox{ otherwise.}
\end{array}\right.\qquad\\
\label{piInverseProps4}
\models w\PiInv(A_1\vee A_2)\equiv(w\PiInv A_1)\vee(w\PiInv A_2)\qquad\\
\label{piInverseProps5}
\models w_1\equiv w_2\mbox{ and }\models A_1\Imp A_2\mbox{ imply }\models w_1\PiInv A_1\Imp w_2\PiInv A_2\qquad
\end{eqnarray}
\end{lemma}
\begin{proof} 
We only do (\ref{piInverseProps3}).
If $\models\Finite\Imp\neg(B\wedge\Box\neg w)$, then $(B\wedge\Box\neg w);\Skip;(A\wedge w)$ is false. Hence applying $w\PiInv.$ gives $\False$ too. An interval $\sigma'$ satisfies $\sigma'\models w\PiInv(B\wedge\Box\neg w;\Skip;(A\wedge w))$ iff it admits the form $\sigma_1'\cdot\sigma_2'$ where $\sigma_1'\models\Finite\wedge B\wedge\Box\neg w$ and $\sigma_2'\models A\wedge w$. Since $\sigma_1'\models\Box\neg w$, $\sigma'|_w=(\sigma_1'\cdot\sigma_2')|_w =\sigma_2'|_w$. Hence $\sigma'|_w\models w\PiInv(A\wedge w)$. Conversely, if $\sigma\models w\PiInv(A\wedge w)$, then $\sigma=\sigma_2'|_w$ for some $\sigma_2'$ such that $\sigma_2'\models A\wedge w$. Then $\sigma=(\sigma_1'\cdot\sigma_2')|_w$ for any $\sigma_1'$ such that $\sigma_1'\models\Finite\wedge B\wedge\Box\neg w$. Hence $\sigma\models(B\wedge\Box\neg w);\Skip;(A\wedge w)$. 
\end{proof}

\qquad 
\begin{proof}[of Theorem \ref{theoremInvPi}]
Consider the system of canonical equations of the form (\ref{keyEqStrictH2}) for $A$. To obtain a system of equations for $w\PiInv A$ from that system, we apply $w\PiInv.$ to all the equations. The result is
\[\begin{array}{lll}
\models w\PiInv(B\wedge w) & \equiv & w\PiInv(B\wedge\Box w)\vee\\
& & 
\bigvee\limits_{k=1}^K w\PiInv((B_k^{ w}\wedge\Box w);\Skip;(B_k^{\neg w,'}\wedge\neg w))\\
[4mm]
\models w\PiInv(B\wedge \neg w) & \equiv & w\PiInv(B\wedge\Box\neg w)\vee\\
& & 
\bigvee\limits_{k=1}^K w\PiInv((B_k^{\neg w}\wedge\Box\neg w);\Skip;(B_k^{w,'}\wedge w))
\end{array}
\]
The validity of these equations follows by the extensionality rule (\ref{piInverseProps5}). By virtue of Remark \ref{remSat}, all the $(B_k^{\varepsilon w}\wedge\Box\varepsilon w)$-s from (\ref{keyEqStrictH2}) can be assumed to be satisfiable. Hence the equivalences (\ref{piInverseProps1})-(\ref{piInverseProps4}) imply that the equations can be simplified to
 \[\begin{array}{l}
\models w\PiInv(B\wedge w)\equiv B\wedge\Box w\vee
\ \bigvee\limits_{k=1}^K (B_k^w\wedge\Box w);\Skip;(w\PiInv(B_k^{\neg w,'}\wedge\neg w))\\
\models w\PiInv(B\wedge \neg w)\equiv 
\ \bigvee\limits_{k=1}^K w\PiInv(B_k^{w,'}\wedge w)
\end{array}
\]
The only designated occurrences of $\PiInv$ in this system of equations are in the unknowns $w\PiInv(B\wedge w)$ and  $w\PiInv(B\wedge \neg w)$, where $B$ ranges over $\Cl^w(A)$ and  over $\Cl^{\neg w}(A)$, respectively. Hence the system admits a $\PiInv$-free solution for all the unknowns, including $w\PiInv(A\wedge w)$ and $w\PiInv (A\wedge \neg w)$. Now $w\PiInv A$ can be expressed using a $\PiInv$-free equivalent to $w\PiInv(A\wedge w)\vee w\PiInv (A\wedge \neg w)$. 
\end{proof}

\qquad

\noindent
The valid equivalences (\ref{piInverseProps1})-(\ref{piInverseProps4}) and the extensionality rule (\ref{piInverseProps5}) from Lemma \ref{lemmaInvPi} can be viewed as a complete axiomatization of $\PiInv$ relative to validity in $\ITL$ because, together with the $\ITL$ reasoning involved in solving the systems of canonical equations from the proof of Theorem \ref{theoremInvPi}, they are sufficient to derive a $\PiInv$-free equivalent formula for every formula in $\ITL$ with $\PiInv$. 

The equivalence (\ref{piInverseProps3}) depends on the satisfiability of a formula of the form $B\wedge\Box w$. This is admissible because validity in $\ITL$ is decidable. Below we argue that such a dependency cannot be avoided. Indeed, if $\PiInv$ is expressible in some conservative extension $\ITL^+$ of $\ITL$, then validity in $\ITL^+$ is either decidable or incomplete. This means that an $\ITL^+$ with $\PiInv$ expressible in it and undecidable validity would admit no recursive axiomatization at all.
\begin{proposition}\label{PiInvIncompleteness}
Let $\ITL^+$ stand for a conservative extension of $\ITL$. Let $A$ be a formula in $\ITL^+$ and let $p\not\in\Var(A)$. Then $\not\models_{\ITL^+}\neg A$ is equivalent to
\begin{equation}\label{incomp1}
\models_{\ITL^++\PiInv}(p\wedge\Empty)\Rightarrow (p\PiInv((p\wedge\Skip);(A\wedge\Box\neg p)))\ .
\end{equation}
Furthermore, $\models_{\ITL^+}A$ is equivalent to 
\begin{equation}\label{incomp2}
\not\models_{\ITL^++\PiInv}\neg(p\wedge\Empty\wedge\neg (p\PiInv((p\wedge\Skip);(\neg A\wedge\Box\neg p))))\ .
\end{equation} 
\end{proposition}
\begin{proof}
(\ref{incomp1}) means that any single state interval $\sigma$ which satisfies $p$ has the form $\sigma'|_p$ where $\sigma'$ is a (possibly right-infinite) interval such that $(\sigma')^0=\sigma^0$ and $(\sigma')^1(\sigma')^2\ldots\models A\wedge\Box\neg p$. Since $p$ does not occur in $A$, if an interval $\sigma''\in(\mathcal{P}(\Var(A)))^+\cup(\mathcal{P}(\Var(A)))^\omega$ that satisfies $A$ exists at all, then $\sigma'\defeq\sigma^0\cdot\sigma''$ has the required property. Conversely, if $\sigma''\in(\mathcal{P}(\Var(A)))^+\cup(\mathcal{P}(\Var(A)))^\omega$ and $\sigma''\models A$, then $(\sigma^0\cdot\sigma'')|_p=\sigma^0$ satisfies $(p\wedge\Empty)\wedge (p\PiInv((p\wedge\Skip);(A\wedge\Box\neg p)))$ for any $\sigma^0\in\mathcal{P}(\Var(A)\cup\{p\})$ such that $p\in \sigma^0$. Intervals $\sigma$ which are not single state, or start with a $\neg p$ state satisfy the first formula trivially.

(\ref{incomp2}) means that a single state interval $\sigma$ which satisfies $p$ and cannot be extended to the right by an interval $\sigma''$ satisfying $\neg A\wedge\Box\neg p$ to obtain a witness interval $\sigma'\defeq\sigma\cdot\sigma''$ for $p\PiInv((p\wedge\Skip);(\neg A\wedge\Box\neg p))$ exists. Since the only restriction imposed on $\sigma''$ is $\sigma''\models \neg A\wedge\Box\neg p$, and $p\not\in\Var(A)$, this means that no interval $\sigma''$ satisfies $\neg A$ at all. Hence $\models_{\ITL^++\PiInv} A$. Conversely, if $\models_{\ITL^++\PiInv} A$, then  $\models_{\ITL^++\PiInv}\neg(p\PiInv((p\wedge\Skip);(\neg A\wedge\Box\neg p)))$. This makes (\ref{incomp2}) equivalent to $\not\models_{\ITL^++\PiInv}\neg(p\wedge\Empty)$, which is true.
\end{proof}

\qquad

\noindent
Proposition \ref{PiInvIncompleteness} shows how the satisfiability of an arbitrary $\ITL^+$ formula can be reduced to the validity of an $\ITL^++\PiInv$ formula, and the validity of an arbitrary $\ITL^+$ formula can be reduced to the satisfiability of an $\ITL^++\PiInv$ formula. If $\PiInv$ is expressible in $\ITL^+$, then the $\ITL^++\PiInv$ formulas in question are each equivalent to some $\ITL^+$ formula. Hence validity and satisfiability are recursively reducible to each other in $\ITL^+$. Then, by Post's theorem, either both $\models_{\ITL^+}$ and $\not\models_{\ITL^+}$ are decidable, or they are both not recursively enumerable. The latter means that if $\models_{\ITL^+}$ is not decidable, then it is not recursively axiomatizable.

\section{Expressing Propositional Quantification in $\ITLNL$}

As stated in the introduction, uniform Craig interpolation can be derived using the expressibility of existential quantification on the non-logical symbols that are not shared between the formulas $A$ and $B$ on the two sides of $\Imp$ in the considered implication $A\Imp B$. Propositional quantification is known to be expressible at the cost of using occurrences of \emph{chop-star} in introspective $\ITL$. This can be demonstrated using that, on finite and right-infinite intervals, $\ITL$ formulas define regular languages with the powerset $\Sigma=\mathcal{P}(V)$ of the vocabulary $V$, which is also the set of states in models, as the alphabet. The classes of regular languages and $\omega$-languages are known to be closed under alphabet homomorphisms, and $\exists p$ is expressed by the homomorphism $h_p:s\mapsto s\setminus\{p\}$. Finally, given the language $L_A$ defined by formula $A$, a regular expression for $h_p(L_A)$ in terms of single-letter languages, set theoretic $\cup$, and formal language operations of concatenation $\cdot$ and Kleene star ${}^*$ and, for the case of $\omega$-languages, Kleene star's `infinite-interval' variant $(.)^\omega$, can be translated back to $\ITL$ by means of $\ITL$'s $\Next$, \emph{chop} and \emph{chop-star}. 

This plan requires a significant change to work for $\ITLNL$. We first need to prove that strictly future $\ITLNL$ formulas $F$, where $\Diamond_r$ can occur along with \emph{chop} and \emph{chop-star}, define regular $\omega$-languages $L_F$ to apply $h_p$ to. A transition from the regular $\omega$-language $h_p(L_F)$ back to the temporal language is possible by a straightforward translation only if infinite intervals are allowed. We do not have infinite intervals in the system of $\ITLNL$ in this paper. Therefore we take the opportunity to use the fact that, with the neighbourhood modalities available, infinite intervals are not essential for defining $\omega$-regular languages in $\ITLNL$. This follows from the availability of the normal form (\ref{ITLNLreactNFConj}) for reactivity. 
\begin{proposition}\label{trFromNLtoITL}
Let $G$ be a Boolean combination of $\Diamond_r$-formulas with future operands. Then there exists a formula $\tr{G}$ in $\ITL$ with right-infinite intervals such that
\begin{equation}\label{trFromNLtoITLEq}
\sigma,0,0\models G\mbox{ iff }\sigma|_{[0,\infty)}\models\tr{G}\ .
\end{equation}
\end{proposition}
\begin{proof}
Assuming the operands of $\Diamond_r$ in $G$ to be in disjunctive normal form, we define the translation $G\mapsto\tr{G}$ from Boolean combinations of $\Diamond_r$-formulas $G$ with future operands to formulas in $\ITL$ with right-infinite intervals by the clauses:
\[\begin{array}{llllll}
\tr{\False} &\defeq &\False  & \tr{\Diamond_r(G_1\vee G_2)} & \defeq & \tr{\Diamond_r G_1}\vee\tr{\Diamond_r G_2}\\
\tr{G_1\Imp G_2} & \defeq & \tr{G_1}\Imp\tr{G_2} & 
\tr{\Diamond_r\left(C\wedge\bigwedge\limits_{n=1}^N\varepsilon_n\Diamond_r G_n\right)} &\defeq &C;\bigwedge\limits_{n=1}^N\tr{\varepsilon_n\Diamond_r G_n}
\end{array}\]
In these clauses, $C$ stands for an introspective formula, and $G$, $G_1$, $G_2$, etc., are future formulas. The case of $N=0$ in the last clause boils down to $\tr{\Diamond_r C}\defeq C;\True$. A plain induction on the construction of $G$ shows that $\tr{.}$ satisfies (\ref{trFromNLtoITLEq}).
\end{proof}

\qquad

\begin{theorem}\label{exprexists}
Let $A$ be an $\ITLNL$ formula and $p$ be a propositional variable. Then $\exists p A$ is equivalent to an $\ITLNL$ formula.
\end{theorem}
\begin{proof}
By separation, we can assume that $A$ is 
\begin{equation}\label{snfA}
\bigvee\limits_k P_k\wedge C_k\wedge F_k
\end{equation}
where $P_k$, $C_k$ and $F_k$ are strictly past, introspective and strictly future, respectively. Then $\exists p\,A$ is equivalent to 
\[\bigvee\limits_k\exists p P_k\wedge\exists p C_k\wedge\exists p F_k.\]
$\exists p$ can be distributed over conjunction as above because, given a reference interval $i,j$ and $\sigma:\Zset\rightarrow\Sigma$ $\sigma,i,j\models P_k$, $\sigma,i,j\models C_k$ and $\sigma,i,j\models F_k$ is determined from $\sigma|_{(-\infty,i-1]}$, $\sigma|_{[i,j]}$ and $\sigma|_{[j+1,\infty)}$, respectively, and these parts of $\sigma$ are disjoint. The existence of $\exists$-free equivalents to $\exists p\,C_k$ is known from the literature because $C_k$ are introspective. It remains to do $\exists p P_k$ and $\exists p F_k$. Using time reversal, we can restrict ourselves to $\exists p F_k$: $\exists p\,P_k$ is equivalent to $(\exists p\,(P_k)^{-1})^{-1}$, and $(P_k)^{-1}$ is a strictly future formula. Hence it is sufficient to show how an $\exists$-free equivalent can be written in $\ITLNL$ for formulas of the form $\exists p\,\Diamond_r(\Skip;F)$ where $F$ is future. Since $\Diamond_r(\Skip;F)$, the general form of strictly future formulas, is equivalent to $\Diamond_r(\Skip\wedge\Diamond_r F)$, Proposition \ref{trFromNLtoITL} implies that 
\[L_{\Diamond_r(\Skip;F)}\defeq \{\sigma|_{[0,\infty)}\in\Sigma^\omega:\sigma,0,0\models \Diamond_r(\Skip;F)\}=\{\sigma\in\Sigma^\omega:\sigma\models G'\}\]
for some formula $G'$ in $\ITL$ with right-infinite intervals. This means that $L_{\Diamond_r(\Skip;F)}$ is regular and so is
\[h_p(L_{\Diamond_r(\Skip;F)})=\{\sigma|_{[0,\infty)}\in\Sigma^\omega:\sigma,0,0\models\exists p\,\Diamond_r(\Skip;F)\}.\]
In Section \ref{reactivityNF} we showed that
\[h_p(L_{\Diamond_r(\Skip;F)})=\{\sigma|_{[0,\infty)}\in\Sigma^\omega:\sigma,0,0\models\bigwedge\limits_{n=1}^N\mathit{Fin}( M_n'')\Imp\mathit{Fin}(M_n')\}\]
where $M_1',M_1'',\ldots,M_N',M_N''$ are some appropriate finite word regular languages and, given an any finite word regular $X\subseteq \Sigma^+$,
\[\mathit{Fin}(X)\defeq \Diamond_r\bigvee\limits_{k=1}^K(C_{X,k}\wedge\Box_r\neg C_{X,k}')\]
where $C_{X,1},C_{X,1}',\ldots,C_{X,K},C_{X,K}'$ satisfy Proposition \ref{keyprop} for $C_X$ being a finite interval $\ITL$ formula which defines $X$. Since finite interval $\ITL$ formulas are nothing but introspective $\ITLNL$ formulas,
\[\models \exists p\,\Diamond_r(\Skip;F)\equiv\bigwedge\limits_{n=1}^N\mathit{Fin}( M_n'')\Imp\mathit{Fin}(M_n').\]
This concludes the proof of the expressibility of $\exists p$ in $\ITLNL$.  
\end{proof}

\section{A Special Case of Propositional Quantification in \emph{chop-star}-free $\ITLNL$}

In this section we use separation to show that propositional quantification with the quantified variable additionally restricted to hold at some bounded number of states is expressible in \emph{chop-star}-free $\ITLNL$.  We use the fact that separation applies in the \emph{chop-star}-free subset of $\ITLNL$. Let
\[\begin{array}{lll}
\sigma,i,j\models\existsMN{n_1}{n_2}p\, A&\mbox{ iff } 
& \mbox{there exists a }\sigma'\sim_p\sigma\mbox{ such that }\sigma',i,j\models A\\
& & \mbox{ and }n_1\leq\#\{k\in\Zset:p\in \sigma^k\}\leq n_2.
\end{array}\]
The technique for eliminating $\existsMN{n_1}{n_2}$ below builds on the expressive completeness proof for $\LTL$ by separation from \cite{GRH94} and the proof of the safety normal form for $\LTL$ in \cite{MP89a}. 
\begin{theorem}\label{starFreeProqQ}
Let $A$ be a \emph{chop-star}-free formula in $\ITLNL$, $p$ be a propositional variable and $0\leq n_1\leq n_2<\omega$. Then there exists a \emph{chop-star}-free $\ITLNL$ formula $A'$ such that $\models A'\equiv\existsMN{n_1}{n_2}p\,A$.
\end{theorem}
Our proof of Theorem \ref{starFreeProqQ} is based on the Lemmata below, which show how $\existsMN{1}{1} p\,B$ can be eliminated for formulas of some particular forms in $\ITLNL$ and finite interval $\ITL$, respectively.
\begin{lemma}\label{starFreeProqQLemma1}
Let $B$ be a \emph{chop-star}-free $\ITLNL$ formula. Then $\existsMN{1}{1} p\,(p\wedge\Empty\wedge B)$ is equivalent to a \emph{chop-star}-free $\ITLNL$ formula.
\end{lemma}
\begin{proof}
Let $\bigvee\limits_k P_k\wedge C_k\wedge F_k$ be a strictly separated equivalent of $B$ in disjunctive normal form with $P_k$, $C_k$ and $F_k$ being strictly past, introspective and strictly future formulas, respectively. Then
\[\models\existsMN{1}{1} p\,(p\wedge\Empty\wedge B)\equiv \Empty\wedge\bigvee\limits_k[\False/p]P_k\wedge [\True/p]C_k\wedge [\False/p]F_k\ .\]
Since $P_k$, $C_k$ and $F_k$ can be assumed to be \emph{chop-star}-free, the above quantifier-free equivalent to $\existsMN{1}{1} p\,(p\wedge\Empty\wedge B)$ is \emph{chop-star}-free too.
\end{proof}
\begin{lemma}\label{starFreeProqQLemma2}
Let $A$ be a \emph{chop-star}-free $\ITL$ formula. Then $\existsMN{1}{1} p\,(p\wedge B)$ is equivalent to a \emph{chop-star}-free $\ITL$ formula.
\end{lemma}
\begin{proof}
Let $B_e\wedge\Empty\vee\bigvee\limits_{k=1}^K B_k\wedge\Next B_k'$ be a GNF of $B$. Then 
\[\models\existsMN{1}{1} p\,(p\wedge B)\equiv [\True/p]B_e\wedge\Empty\vee\bigvee\limits_{k=1}^K [\True/p]B_k\wedge\Next [\False/p]B_k'\ .\]
Since $B_k'$ can be assumed to be \emph{chop-star}-free, the above quantifier-free equivalent to $\existsMN{1}{1} p\,(p\wedge B)$ is \emph{chop-star}-free too.
\end{proof}

\qquad

\begin{proof}[of Theorem \ref{starFreeProqQ}]
The case of $n_1<n_2$ reduces to the case $n_1=n_2$ by induction on $n_1$ using 
\[\models \existsMN{n_1}{n_2}p\,A\equiv \existsMN{n_1}{n_1}p\,A\vee \existsMN{n_1+1}{n_2}p\,A.\]
The case $n\defeq n_1=n_2$ reduces to the case $n=n_1=n_2=1$ by induction on $n$ using
\[\models\existsMN{0}{0}p\,A\equiv[\False/p]A\ \mbox{ and }\models\existsMN{n+1}{n+1}p\,A\equiv \existsMN{n}{n}p\,\existsMN{1}{1}q\,(\Box_a\neg(p\wedge q)\wedge[p\vee q/p]A)\]
where $q$ is a fresh propositional variable.
Hence it remains to do the case $\existsMN{1}{1}q\,B$. Let $r_1$ and $r_2$ be two more fresh variables. We have
\[\models\existsMN{1}{1}q\,B\equiv\existsMN{1}{1}r_1\existsMN{1}{1}r_2(r_1\wedge\Fin r_2\wedge\Diamond_a\existsMN{1}{1}q(q\wedge\Empty\wedge\Diamond_a(r_1\wedge\Fin r_2\wedge B)))\ .\]
In words, $\Diamond_a\existsMN{1}{1}q(q\wedge\Empty\wedge\Diamond_a(r_1\wedge\Fin r_2\wedge B))$ moves from the reference interval to the unique single state interval which, according to $\existsMN{1}{1}q$, satisfies $q$, and $r_1$ and $r_2$ mark the endpoints of the reference interval to ensure that $\Diamond_a(r_1\wedge\Fin r_2\wedge B)$ means that $B$ is evaluated at the original reference interval.

By Lemma \ref{starFreeProqQLemma1}, $\existsMN{1}{1}q(q\wedge\Empty\wedge\Diamond_a(r_1\wedge\Fin r_2\wedge B))$ is equivalent to some \emph{chop-star}-free $\ITLNL$ formula $G'$. Hence 
\[\models\existsMN{1}{1}q\,B\equiv\existsMN{1}{1}r_1\existsMN{1}{1}r_2(r_1\wedge\Fin r_2\wedge\Diamond_a G').\]
Let $\bigvee\limits_{k=1}^K P_k\wedge C_k\wedge F_k$ be a \emph{chop-star}-free strictly separated equivalent of $\Diamond_a G'$ in disjunctive normal form, with $P_k$, $C_k$ and $F_k$ being strictly past, introspective and strictly future, respectively. Then
\[\models\existsMN{1}{1}q\,B\equiv\bigvee\limits_{k=1}^K[\False/r_1,\False/r_2]P_k\wedge \existsMN{1}{1}r_1\existsMN{1}{1}r_2(r_1\wedge \Fin r_2\wedge C_k)\wedge [\False/r_1,\False/r_2]F_k\ .\]
To eliminate $\existsMN{1}{1}$ from $\existsMN{1}{1}r_1\existsMN{1}{1}r_2(r_1\wedge \Fin r_2\wedge C_k)$ we use that
\[\existsMN{1}{1}r_1\existsMN{1}{1}r_2(r_1\wedge \Fin r_2\wedge C_k)\equiv\existsMN{1}{1}r_1(r_1\wedge (\existsMN{1}{1}r_2(r_2\wedge (C_k)^{-1}))^{-1})\ ,\]
where $(.)^{-1}$ is time reversal. By Lemma \ref{starFreeProqQLemma2}, $\existsMN{1}{1}r_2(r_2\wedge (C_k)^{-1})$ has some quantifier-free and \emph{chop-star}-free equivalent $G_k'$ and $\existsMN{1}{1}r_1(r_1\wedge (\existsMN{1}{1}r_2(r_2\wedge (C_k)^{-1}))^{-1})$, which is therefore equivalent to $\existsMN{1}{1}r_1(r_1\wedge (G_k')^{-1})$, has a quantifier-free and \emph{chop-star}-free equivalent too. This concludes the elimination of $\existsMN{1}{1}p$ and, consequently, $\existsMN{n}{n} p$ for all natural $n$, and $\existsMN{n_1}{n_2} p$ for all $n_1\leq n_2$.
\end{proof}

\qquad

\noindent
Theorem \ref{starFreeProqQ} does not apply in case the number of states where $p$ holds is restricted only from below because, e.g.,
\[\models\existsMN {n}{\omega}p\bigl(\underbrace{(p\wedge\Skip);\ldots;(p\wedge\Skip)}_{n\mbox{\scriptsize \ times}};A\bigr)\equiv\underbrace{\Skip;\ldots;\Skip}_{n\mbox{\scriptsize \ times}};\exists p\,A.\]

\section{Craig Interpolation and Beth Definability for $\ITLNL$}

Craig interpolation for $\ITLNL$ follows almost immediately from Theorem \ref{exprexists}.
\begin{corollary}[Strongest Consequence in $\ITLNL$]
Let $A$ be a formula in $\ITLNL$ and $p_1,\ldots,p_N$ be some propositional variables. Then there exists an $\ITLNL$ formula $C$ such that $\Var(C)\subseteq\Var(A)\setminus\{p_1,\ldots,p_N\}$ and $\models A\Imp B$ is equivalent to $\models C\Imp B$ for every $\ITLNL$ formula $B$ such that $\Var(B)\subseteq\Var(A)\setminus\{p_1,\ldots,p_N\}$.
\end{corollary}
A formula satisfying the above requirement on $C$ is called the {\em strongest consequence of $A$ in terms of $\Var(A)\setminus\{p_1,\ldots,p_N\}$}.

\noindent 
\begin{proof} 
$C$ can be chosen to be an $\exists$-free equivalent to $\exists p_1\ldots \exists p_N A$ such as what is guaranteed to exist by Theorem \ref{exprexists}.
\end{proof}
\begin{corollary}[Uniform Craig Interpolation for $\ITLNL$]
Given $\ITLNL$ formulas $A$ and $B$, if $\models A\Imp B$, then there exists a formula $C$ such that $\Var(C)\subseteq\Var(A)\cap \Var(B)$ and $\models A\Imp C$, $\models C\Imp B$. $C$ can be determined from $A$ only, and satisfy the requirement for all $B$ which produce the same $\Var(A)\cap\Var(B)$.
\end{corollary}
\begin{proof} 
$C$ can be chosen to be any strongest consequence of $A$ in terms of $\Var(A)\cap\Var(B)$.
\end{proof}

\qquad

\noindent
In (\ref{bethLTL}) $\Box$ works as the {\em universal modality} in (future only) $\LTL$.  
In $\ITLNL$, the universal modality is $\Box_a$. Using $\Box_a$, {\em $\ITLNL$ formula $A$ implicitly defines propositional variable $p$} is written as:
\begin{equation}\label{bethITLNL}
\models \Box_a A\wedge \Box_a[p'/p]A\Imp\Box_a(p\equiv p')\ .
\end{equation}
\begin{corollary}[Beth Definability for $\ITLNL$]
Given $\ITLNL$ formula $A$ such that (\ref{bethITLNL}) holds, there exists a formula $C$ such that $\Var(C)\subseteq\Var(A)\setminus\{p\}$ and 
\[\models \Box_a A\Imp \Box_a(p\equiv C).\]
\end{corollary}
\begin{proof}
Since $\models\Box_a A\wedge \Box_a[p'/p]A\Imp\Box_a(p\equiv p')$ and $\models\Box_a A\wedge \Box_a[p'/p]A\Imp(p\equiv p')$ are equivalent, we can apply interpolation to $\models(\Box_a A\wedge p)\Imp(\Box_a[p'/p]A\Imp p')$ and obtain a $p$- and $p'$-free $C$ such that $\models(\Box_a A\wedge p)\Imp C$ and $\models C\Imp(\Box_a[p'/p]A\Imp p')$. The latter is equivalent to $\models C\Imp(\Box_a A\Imp p)$. Since  $\models(\Box_a A\wedge p)\Imp C$ and $\models C\Imp(\Box_a A\Imp p)$ are equivalent to $\models\Box_a A\Imp(p\Imp C)$ and $\models\Box_a\Imp (C\Imp p)$, respectively, we obtain $\models\Box_a A\Imp (p\equiv C)$, just like in the archetype case of predicate logic.  
\end{proof}

\section*{Concluding Remarks}

We have shown how interval-based temporal separation in $\ITLNL$ and a lemma which is key for establishing this form of separation in \cite{GM2022} can be used to establish a number of results on $\ITL$ and $\ITLNL$ by concise proofs. These include an interval-based variant of the reactivity normal form as known from \cite{MP89a}, the expressibility of the inverse of the temporal projection operator from \cite{DBLP:conf/icalp/HalpernMM83}, a normal form for $\ITL$ formulas $A$ which, given a state formula $w$, features the conditions that the maximal $\Box w$- and $\Box\neg w$-subintervals of an interval satisfying $A$ need to satisfy, the elimination of propositional quantification in $\ITLNL$ and, as corollaries, uniform Craig interpolation and Beth definability for $\ITLNL$, and the elimination of propositional quantification with an upper bound on the number of states which satisfy the quantified variable in the \emph{chop-star}-free subset of $\ITLNL$.

Notably, the $\LTL$-based forms of these results, except that on $\PiInv$, which is a new connective in $\ITL$, along with separation on its own, are well-understood and have shown to be very useful in the literature. Uniform Craig interpolation and Beth definability are an exception; $\LTL$ does not have them as pointed out in \cite{Maksimova91}. They hold in $\ITL$ (without the neighbourhood modalities) too. However, the availability of the neighbourhood modalities, which make it possible to use $\Diamond_a\defeq\Diamond_r\Diamond_r\Diamond_l\Diamond_l$ as the universal modality, greatly enhance the scope of explicit definitions $\Box_a(p\equiv C)$: the restriction on the semantics of atomic propositions $p$ imposed by the locality principle in $\ITL$ can be largely compensated for by letting $\Diamond_l$ and $\Diamond_r$ occur in interpolants $C$.

\section{Acknowledgements} 
The author thanks Antonio Cau and Ben Moszkowski for some stimulating discussions and comments on this work.

\bibliographystyle{alpha}
\bibliography{separationInITL}

@Misc{CMZ,
  author = "Antonio Cau and Ben Moszkowski and Hussein Zedan",
  title = "{ITL web pages}",
  howpublished = "URL: {\tt http://www.antonio-cau.co.uk/ITL/}"
}

@article{Mos85,
  author = "Ben Moszkowski",
  title = "{Temporal Logic For Multilevel Reasoning About Hardware}",
  journal = "IEEE Computer",
  year = "1985",
  volume = "18",
  number = "2",
  pages = "10-19"
}

@book{Mos86,
  author = "Ben Moszkowski",
  title = "{Executing Temporal Logic Programs}",
  year = "1986",
  publisher = "Cambridge University Press",
  note = "{URL: {\tt http://www.antonio-cau.co.uk/ITL/publications/reports/tempura-book.pdf}}"
}

@article{ZHR91,
  author = "{Zhou Chaochen} and C. A. R. Hoare and Anders P. Ravn",
  title = "{A Calculus of Durations}",
  journal = "Information Processing Letters",
  year = "1991",
  volume = "40",
  number = "5",
  pages = "269-276"
}

@article{DBLP:journals/eatcs/MonicaGMS11,
  author    = {Dario Della Monica and
               Valentin Goranko and
               Angelo Montanari and
               Guido Sciavicco},
  title     = {{Interval Temporal Logics: a Journey}},
  journal   = {Bull. {EATCS}},
  volume    = {105},
  pages     = {73--99},
  year      = {2011},
  bibsource = {dblp computer science bibliography, https://dblp.org}
}

@article{DBLP:journals/jancl/GorankoMS04,
  author    = {Valentin Goranko and
               Angelo Montanari and
               Guido Sciavicco},
  title     = {{A Road Map of Interval Temporal Logics and Duration Calculi}},
  journal   = {J. Appl. Non Class. Logics},
  volume    = {14},
  number    = {1-2},
  pages     = {9--54},
  year      = {2004},
  doi       = {10.3166/jancl.14.9-54},
  bibsource = {dblp computer science bibliography, https://dblp.org}
}

@inproceedings{Pnu77,
  author = "Amir Pnueli",
  title = "{The Temporal Logic of Programs}",
  year = "1977",
  booktitle = "{Proceedings of the 18th IEEE Symposium Foundations of Computer Science}",
  pages = "46-57",
  publisher = "IEEE"
}

@inproceedings{Gab89,
  author = "Dov M. Gabbay",
  title = "{Declarative Past and Imperative Future: Executable Temporal Logic for Interactive Systems}",
  booktitle = "{Proceedings of the Colloquium of Temporal Logic in Specification}",
  pages = "67-89",
  year = "1989",
  volume = "398",
  series = "LNCS",
  publisher = "Springer"
}

@book{GRH94,
   author       = "Dov Gabbay and  Ian Hodkinson and Mark Reynolds",
   title        = "{Temporal Logic: Mathematical Foundations and Computational Aspects. Volume I}",
   publisher    = {Oxford University Press},
   year         = 1994
}

@inproceedings{DBLP:conf/csl/Rasmussen99,
  author    = {Thomas Mathredal Rasmussen},
  editor    = {J{\"{o}}rg Flum and
               Mario Rodr{\'{\i}}guez{-}Artalejo},
  title     = {{Signed Interval Logic}},
  booktitle = {{CSL} '99, Proceedings},
  series    = {LNCS},
  volume    = {1683},
  pages     = {157--171},
  publisher = {Springer},
  year      = {1999},
  url       = {https://doi.org/10.1007/3-540-48168-0\_12},
  doi       = {10.1007/3-540-48168-0\_12},
  bibsource = {dblp computer science bibliography, https://dblp.org}
}

@article{DBLP:journals/iandc/Rabinovich00,
  author    = {Alexander Moshe Rabinovich},
  title     = {{Expressive Completeness of Duration Calculus}},
  journal   = {Inf. Comput.},
  volume    = {156},
  number    = {1-2},
  pages     = {320--344},
  year      = {2000},
  url       = {https://doi.org/10.1006/inco.1999.2816},
  doi       = {10.1006/inco.1999.2816},
  bibsource = {dblp computer science bibliography, https://dblp.org}
}

@inproceedings{HS86,
  author = "J. Y. Halpern and Y. Shoham",
  title = "{A Propositional Logic of Time Intervals}",
  booktitle = "{Proceedings of LICS'86}",
  year = "1986",
  pages = "279-292",
  publisher = "IEEE Computer Society Press"
}

@article{DBLP:journals/cacm/Allen83,
  author    = {James F. Allen},
  title     = {{Maintaining Knowledge about Temporal Intervals}},
  journal   = {Commun. {ACM}},
  volume    = {26},
  number    = {11},
  pages     = {832--843},
  year      = {1983},
  doi       = {10.1145/182.358434},
  bibsource = {dblp computer science bibliography, https://dblp.org}
}

@inproceedings{MP89a,
  author = "Zohar Manna and Amir Pnueli",
  title = "{A Hierarchy of Temporal Properties}",
  booktitle = "{9th Symposium on Principles of Distributed Computing}",
  pages =        "377-408",
  year =         "1990",
  publisher = "ACM Press"
}

@inproceedings{DBLP:conf/dagstuhl/Farwer01,
  author    = {Berndt Farwer},
  editor    = {Erich Gr{\"{a}}del and
               Wolfgang Thomas and
               Thomas Wilke},
  title     = {$\omega$-{Automata}},
  booktitle = {Automata, Logics, and Infinite Games: {A} Guide to Current Research},
  series    = {LNCS},
  volume    = {2500},
  pages     = {3--20},
  publisher = {Springer},
  year      = {2001},
  url       = {https://doi.org/10.1007/3-540-36387-4\_1},
  doi       = {10.1007/3-540-36387-4\_1},
  bibsource = {dblp computer science bibliography, https://dblp.org}
}

@phdthesis{Zuc86,
  author = "Lenore Zuck",
  title = "{Past Temporal Logic}",
  school = "Weizmann Institute of Science",
  year = "1986",
  type = "{Ph.D. thesis}"
}

@article{DBLP:journals/logcom/Venema91,
  author    = {Yde Venema},
  title     = {{A Modal Logic for Chopping Intervals}},
  journal   = {J. Log. Comput.},
  volume    = {1},
  number    = {4},
  pages     = {453--476},
  year      = {1991},
  doi       = {10.1093/logcom/1.4.453},
  bibsource = {dblp computer science bibliography, https://dblp.org}
}

@inproceedings{DBLP:conf/compos/ChaochenH97,
  author    = {{Zhou Chaochen} and
               Michael R. Hansen},
  editor    = {Willem P. de Roever and
               Hans Langmaack and
               Amir Pnueli},
  title     = {{An Adequate First Order Interval Logic}},
  booktitle = {COMPOS'97. Revised Lectures},
  series    = {LNCS},
  volume    = {1536},
  pages     = {584--608},
  publisher = {Springer},
  year      = {1997},
  url       = {https://doi.org/10.1007/3-540-49213-5\_23},
  doi       = {10.1007/3-540-49213-5\_23},
  bibsource = {dblp computer science bibliography, https://dblp.org}
}

@inproceedings{DBLP:conf/icalp/HalpernMM83,
  author    = {Joseph Y. Halpern and
               Zohar Manna and
               Ben C. Moszkowski},
  editor    = {Josep D{\'{\i}}az},
  title     = {{A Hardware Semantics Based on Temporal Intervals}},
  booktitle = {ICALP 1983, Proceedings},
  series    = {LNCS},
  volume    = {154},
  pages     = {278--291},
  publisher = {Springer},
  year      = {1983},
  url       = {https://doi.org/10.1007/BFb0036915},
  doi       = {10.1007/BFb0036915},
  bibsource = {dblp computer science bibliography, https://dblp.org}
}

@article{GM2022,
  author    = {Dimitar P. Guelev and
               Ben C. Moszkowski},
  title     = {{A Separation Theorem for Discrete-time Interval Temporal Logic}},
  journal   = {J. Appl. Non Class. Logics},
  volume    = {32},
  number    = {1},
  pages     = {28--54},
  year      = {2022},
  doi       = {10.1080/11663081.2022.2050135},
  bibsource = {dblp computer science bibliography, https://dblp.org}
}

@article{GM2024,
title = {{Expressive Completeness by Separation for Discrete Time Interval Temporal Logic with Expanding Modalities}},
journal = {Information Processing Letters},
volume = {186},
pages = {106480},
year = {2024},
issn = {0020-0190},
doi = {https://doi.org/10.1016/j.ipl.2024.106480},
url = {https://www.sciencedirect.com/science/article/pii/S0020019024000103},
author = {Dimitar P. Guelev and Ben Moszkowski}
}

@phdthesis{Kamp68,
  author = "J. A. W. Kamp",
  title = "{Tense Logic and the Theory of Linear Order}",
  school = "University of California",
  year = "1968",
  type = "{Ph.D. thesis}",
  address = "Los Angeles"
}

@phdthesis{MosThesis,
  author = "Benjamin C. Moszkowski",
  title = "{Reasoning about Digital Circuits}",
  school = "Stanford University",
  year = "1983",
  type = "{Ph.D. thesis}",
  address = "Stanford"
}

@book{hughes1996new,
  title="{A New Introduction to Modal Logic}",
  author={Hughes, G.E. and Cresswell, M.J.},
  isbn={9780415125994},
  lccn={95014728},
  url={https://books.google.bg/books?id=Dsn1xWNB4MEC},
  year={1996},
  publisher={Routledge}
}

@article{DBLP:journals/logcom/BaruaRC00,
  author       = {Rana Barua and
                  Suman Roy and
                  {Zhou Chaochen}},
  title        = "{Completeness of neighbourhood logic}",
  journal      = {J. Log. Comput.},
  volume       = {10},
  number       = {2},
  pages        = {271--295},
  year         = {2000},
  url          = {https://doi.org/10.1093/logcom/10.2.271},
  doi          = {10.1093/LOGCOM/10.2.271},
  bibsource    = {dblp computer science bibliography, https://dblp.org}
}

@book{DBLP:series/eatcs/ChaochenH04,
  author       = {{Zhou Chaochen} and
                  Michael R. Hansen},
  title        = "{Duration Calculus - {A} Formal Approach to Real-Time Systems}",
  series       = {Monographs in Theoretical Computer Science. An {EATCS} Series},
  publisher    = {Springer},
  year         = {2004},
  url          = {https://doi.org/10.1007/978-3-662-06784-0},
  doi          = {10.1007/978-3-662-06784-0},
  isbn         = {978-3-642-07404-2},
  bibsource    = {dblp computer science bibliography, https://dblp.org}
}

@article{Maksimova91,
  author       = {Larissa L. Maksimova},
  title        = {{Absence of interpolation and of Beth’s property in temporal logics with “the
next” operation}},
  journal      = {Siberian Mathematical Journal},
  volume       = {32},
  number       = {6},
  pages        = {989 - 993},
  year         = {1991}
}

@inproceedings{DBLP:conf/csl/GheerbrantC09,
  author       = {Am{\'{e}}lie Gheerbrant and
                  Balder ten Cate},
  editor       = {Erich Gr{\"{a}}del and
                  Reinhard Kahle},
  title        = {{Craig Interpolation for Linear Temporal Languages}},
  booktitle    = {Proceedings of {CSL} 2009},
  series       = {LNCS},
  volume       = {5771},
  pages        = {287--301},
  publisher    = {Springer},
  year         = {2009},
  url          = {https://doi.org/10.1007/978-3-642-04027-6\_22},
  doi          = {10.1007/978-3-642-04027-6\_22},
  bibsource    = {dblp computer science bibliography, https://dblp.org}
}

@book{Sho67,
  author = "Joseph Shoenfield",
  title = "{Mathematical Logic}",
  year = "1967",
  address = "Reading, Massachusetts",
  month = "",
  publisher = "Addison-Wesley"
}

@article{GorankoPassy1992,
    author = {Valentin Goranko and Solomon Passy},
    title = {{Using the Universal Modality: Gains and Questions}},
    journal = {Journal of Logic and Computation},
    volume = {2},
    number = {1},
    pages = {5-30},
    year = {1992},
    month = {03},
    issn = {0955-792X},
    doi = {10.1093/logcom/2.1.5},
    url = {https://doi.org/10.1093/logcom/2.1.5},
    eprint = {https://academic.oup.com/logcom/article-pdf/2/1/5/2990585/2-1-5.pdf},
}

@inproceedings{DBLP:conf/icalp/EisnerFHMC03,
  author       = {Cindy Eisner and
                  Dana Fisman and
                  John Havlicek and
                  Anthony McIsaac and
                  David Van Campenhout},
  editor       = {Jos C. M. Baeten and
                  Jan Karel Lenstra and
                  Joachim Parrow and
                  Gerhard J. Woeginger},
  title        = {{The Definition of a Temporal Clock Operator}},
  booktitle    = {Proceedings of {ICALP} 2003},
  series       = {LNCS},
  volume       = {2719},
  pages        = {857--870},
  publisher    = {Springer},
  year         = {2003},
  url          = {https://doi.org/10.1007/3-540-45061-0\_67},
  doi          = {10.1007/3-540-45061-0\_67},
  bibsource    = {dblp computer science bibliography, https://dblp.org}
}

@article{Koz83,
  author = "Dexter Kozen",
  title = "{Results on the propositional $\mu$-calculus}",
  journal = "Theoretical Computer Science",
  year = "1983",
  volume = "27",
  pages = "333-354"
}

@inproceedings{DBLP:conf/time/Guelev22,
  author       = {Dimitar P. Guelev},
  editor       = {Alexander Artikis and
                  Roberto Posenato and
                  Stefano Tonetta},
  title        = {{Gabbay Separation for the Duration Calculus}},
  booktitle    = {Proceedings of {TIME} 2022},
  series       = {LIPIcs},
  volume       = {247},
  pages        = {10:1--10:14},
  publisher    = {Schloss Dagstuhl - Leibniz-Zentrum f{\"{u}}r Informatik},
  year         = {2022},
  url          = {https://doi.org/10.4230/LIPIcs.TIME.2022.10},
  doi          = {10.4230/LIPICS.TIME.2022.10},
  bibsource    = {dblp computer science bibliography, https://dblp.org}
}

@article{DBLP:journals/iandc/Wagner79,
  author       = {Klaus W. Wagner},
  title        = {{On omega-Regular Sets}},
  journal      = {Inf. Control.},
  volume       = {43},
  number       = {2},
  pages        = {123--177},
  year         = {1979},
  url          = {https://doi.org/10.1016/S0019-9958(79)90653-3},
  doi          = {10.1016/S0019-9958(79)90653-3},
  bibsource    = {dblp computer science bibliography, https://dblp.org}
}

@article{DBLP:journals/mst/Landweber69,
  author       = {Lawrence H. Landweber},
  title        = {{Decision Problems for omega-Automata}},
  journal      = {Math. Syst. Theory},
  volume       = {3},
  number       = {4},
  pages        = {376--384},
  year         = {1969},
  url          = {https://doi.org/10.1007/BF01691063},
  doi          = {10.1007/BF01691063},
  bibsource    = {dblp computer science bibliography, https://dblp.org}
}

@InProceedings{SafetyProgress10.1007/978-3-642-58041-3_5,
author="Chang, Edward
and Manna, Zohar
and Pnueli, Amir",
editor="Bauer, Friedrich L.
and Brauer, Wilfried
and Schwichtenberg, Helmut",
title="{The Safety-Progress Classification}",
booktitle="Logic and Algebra of Specification",
year="1993",
publisher="Springer Berlin Heidelberg",
address="Berlin, Heidelberg",
pages="143--202",
isbn="978-3-642-58041-3"
}

@inproceedings{DBLP:conf/concur/Kaivola95,
  author       = {Roope Kaivola},
  editor       = {Insup Lee and
                  Scott A. Smolka},
  title        = {{Axiomatising Linear Time Mu-calculus}},
  booktitle    = {Proceedings of {CONCUR} '95},
  series       = {LNCS},
  volume       = {962},
  pages        = {423--437},
  publisher    = {Springer},
  year         = {1995},
  url          = {https://doi.org/10.1007/3-540-60218-6\_32},
  doi          = {10.1007/3-540-60218-6\_32},
  bibsource    = {dblp computer science bibliography, https://dblp.org}
}

@incollection{DBLP:books/el/leeuwen90/Thomas90,
  author       = {Wolfgang Thomas},
  editor       = {Jan van Leeuwen},
  title        = {{Automata on Infinite Objects}},
  booktitle    = {Handbook of Theoretical Computer Science, Volume {B:} Formal Models
                  and Semantics},
  pages        = {133--191},
  publisher    = {Elsevier and {MIT} Press},
  year         = {1990},
  url          = {https://doi.org/10.1016/b978-0-444-88074-1.50009-3},
  doi          = {10.1016/B978-0-444-88074-1.50009-3},
  bibsource    = {dblp computer science bibliography, https://dblp.org}
}

@article{DBLP:journals/fac/MoszkowskiG17,
  author       = {Ben C. Moszkowski and
                  Dimitar P. Guelev},
  title        = {{An Application of Temporal Projection to Interleaving Concurrency}},
  journal      = {Formal Aspects Comput.},
  volume       = {29},
  number       = {4},
  pages        = {705--750},
  year         = {2017},
  url          = {https://doi.org/10.1007/s00165-017-0417-3},
  doi          = {10.1007/S00165-017-0417-3},
  bibsource    = {dblp computer science bibliography, https://dblp.org}
}

@phdthesis{DBLP:phd/ethos/Duan96,
  author       = {Zhenhua Duan},
  title        = {{An Extended Interval Temporal Logic and a Framing Technique for Temporal
                  Logic Programming}},
  school       = {Newcastle University, Newcastle upon Tyne, {UK}},
  year         = {1996},
  url          = {https://hdl.handle.net/10443/2075},
  bibsource    = {dblp computer science bibliography, https://dblp.org}
}

@article{DBLP:journals/fac/ZhangMZL24,
  author       = {Yuanrui Zhang and
                  Fr{\'{e}}d{\'{e}}ric Mallet and
                  Min Zhang and
                  Zhiming Liu},
  title        = {{Specification and Verification of Multi-Clock Systems Using a Temporal
                  Logic with Clock Constraints}},
  journal      = {Formal Aspects Comput.},
  volume       = {36},
  number       = {2},
  pages        = {13},
  year         = {2024},
  url          = {https://doi.org/10.1145/3670794},
  doi          = {10.1145/3670794},
  bibsource    = {dblp computer science bibliography, https://dblp.org}
}

@inproceedings{PD97,
  author = "Paritosh K. Pandya and {Dang Van Hung}",
  title = "{Duration Calculus of Weakly Monotonic Time}",
  booktitle = "{Proceedings of FTRTFT'98}",

  year = "1998",
  volume = "1486",
  series = "LNCS",
  pages = "55-64",
  publisher = "Springer"
}

\end{document}